\documentclass[11pt]{iopart}
\usepackage{epsfig,psfrag}
\usepackage{iopams}
\usepackage{amsfonts}
\usepackage{mathrsfs}
\usepackage{bm}
\usepackage{color}
\usepackage{amssymb}
\newcommand\beq{\begin{equation}}
\newcommand\eeq{\end{equation}}
\newcommand\beqa{\begin{eqnarray}}
\newcommand\eeqa{\end{eqnarray}}

\def\half{\frac{1}{2}}

\def\eps{\epsilon}

\newcommand\bx{\mathbf{x}}
\newcommand\bz{\mathbf{\hat{z}}}
\newcommand\bX{\mathbf{X}}
\newcommand\bff{\mathbf{f}}
\newcommand\bF{\mathbf{F}}

\newcommand\bu{\mathbf{u}}

\newcommand {\pdd}[2]{\frac{\partial #1}{\partial #2}}

\newcommand {\vel}{\mathbf{u}}

\newcommand {\avel}{\bnu}
\newcommand {\La}{{\cal L}}

\begin{document}

\review[]{An optimisation approach for analysing nonlinear stability with transition to turbulence in fluids as an exemplar} 

\author{R R Kerswell$^1$, C C T Pringle$^2$ and A P Willis$^3$}
\address{$^1$ School of Mathematics, University of Bristol, Bristol BS8 1TW, UK}
\address{$^2$ Applied Mathematics Research Centre, Faculty of Engineering and Computing, Coventry University, Coventry CV1 5FB, UK}
\address{$^3$ School of Mathematics and Statistics, University of Sheffield, Sheffield S3 7RH, UK}

\ead{R.R.Kerswell@bris.ac.uk}

\vspace{1cm}
\noindent 
\hspace{2.5cm} {submitted 13th February 2014}\\

\begin{abstract}\\

This article introduces, and reviews recent work using, a simple
optimisation technique for analysing the nonlinear stability of a
state in a dynamical system. The technique can be used to identify the
most efficient way to disturb a system such that it transits from one
stable state to another. The key idea is introduced within the
framework of a finite-dimensional set of ordinary differential
equations (ODEs) and then illustrated for a very simple system of 2
ODEs which possesses bistability. Then the transition to turbulence
problem in fluid mechanics is used to show how the technique can be
formulated for a spatially-extended system described by a partial
differential equation (the well-known Navier-Stokes equation). Within
that context, the optimisation technique bridges the gap between
(linear) optimal perturbation theory and the (nonlinear) dynamical
systems approach to fluid flows. The fact that the technique has now
been recently shown to work in this very high dimensional setting
augurs well for its utility in other physical systems.

\end{abstract}


\maketitle 

\section{Introduction}

Many physical systems possess a multiplicity of stable states so that
more than one solution or system configuration can be found at long
times. In such situations, a key issue is usually maintaining the
system in a desired state against ambient noise or switching the
system from one (undesirable) state to another (preferred) state in an
efficient, robust way: e.g. in liquid crystal displays \cite{KG02},
power grids \cite{MBB08}, arrays of coupled lasers \cite{E05},
turbulent fluid flows \cite{G00} and even in the human brain
\cite{BD86}. Either objective involves detailed knowledge of a state's
basin of attraction, defined as the set of all initial conditions of
the system whose long time behaviour is to converge to that
state. Initial conditions located just outside the basin boundary
indicate how the system can be efficiently disturbed to trigger a new
stable state. Knowledge of how the basin boundary of a state moves (in
phase space) when the system is manipulated (e.g. by modifying the
boundary conditions) opens up the possibility of enhancing the
nonlinear stability of that state to finite amplitude
disturbances. However, locating a basin boundary is a fully nonlinear
(nonlocal) problem so that the traditional tools of linearising the
system around the state or even weakly nonlinear analysis provide no
traction. Existing fully nonlinear approaches - solving the governing
equations while searching for the finite-amplitude disturbances to
just knock the system out of one state into another, or mapping out
the stable and unstable manifolds of nearby solutions in phase space
to identify the basin boundary - are impractical for all but the
smallest systems.

Recently, a new, very general, fully nonlinear optimisation technique
has emerged as a viable way to make progress. The underlying idea is
relatively simple and, perhaps because of this, seems to have been
formulated independently in (at least) three different parts of the
scientific literature over the last decade (transitional shear flows
\cite{CDRB10,M11,PK10} - see \S \ref{transition}, oceanography \cite{MSD04} 
- see \S \ref{ocean} and thermoacoustics \cite{J11} - see \S \ref{thermo}).  
The key advance, however, has come in the
last few years when the feasibility of the approach has been
demonstrated for the 3-dimensional Navier-Stokes equations discretised
by a large number ($O(10^5$-$10^6)$) of degrees of freedom
\cite{CDRB10,CDRB11,CD13,D13,M11,PK10,PWK12,RCK12}. This suggests that 
other partial differential equation systems could be
usefully analysed with this approach.

This article (which is an updated and extended version of \cite{K11})
is an attempt to provide a simple introduction to the idea and to
review the progress made so far. As should become clear, the approach
is still developing but there is already enough evidence garnered to
indicate that it adds something quite new to a theoretician's
toolbox. In fluid mechanics, the last two decades have seen a huge
amount of work looking at flow transition either from the linear
transient growth perspective (also called `non-modal analysis' or
`optimal perturbation theory' \cite{Gr00,R01,SH01,S07}) or more
recently in terms of exact solutions and their manifolds (a dynamical
systems approach \cite{K05,E07,KUV12}). The optimisation technique
discussed here bridges the well-known `amplitude' gap between these
two viewpoints by extending the (infinitesimal amplitude) transient
growth optimal of the former approach into finite amplitudes and
ultimately up to where the basin boundary is crossed (the closest
stable manifold of a nearby exact solution).
 
The plan of this article is as follows. In section 2, the central idea
of the optimisation approach is introduced in the context of a
finite-dimensional dynamical system.  A simple system of 2 ordinary
differential equations (ODEs) is then used to: a) illustrate the
results of a calculation, and b) highlight some important ingredients
which make the technique work.  In section 3, the discussion is moved
onto fluid mechanics and the Navier-Stokes equation - a
time-dependent and 3-space dimension nonlinear partial differential
equation (PDE). The first attempted use of the optimisation technique
was in the classical problem of flow through a pipe \cite{PK10} and so
this is used here as the context to explain the inner workings of the
approach. Appendices A and B provide simple supporting illustrations
of a linear transient growth calculation for a system of 2 ODEs and how
including nonlinearity can allow different growth mechanisms to work
together to produce far greater overall growth. Section 4 reviews some
of the literature which preceded the successful application to the
Navier-Stokes equations as well as discussing further work now
building on it. Finally section 5 provides a summary and surveys what
the future may hold.

\section{Basic Idea}

To explain the optimisation technique, consider a finite-dimensional dynamical
system 
\beq
\frac{d\bX}{dt}=\bff(\bX;\mu)
\eeq
where $\bX=\bX(t) \in {\mathbb R}^N$ and $\mu$ is
a parameter of the system. Let $\bX_0$ be a local steady attractor of
interest and $\bx:=\bX-\bX_0$ be the perturbation away from this
attractor. Then write the evolution equation for $\bx$ as
\begin{equation}
\frac{d\bx}{dt}=\bF(\bx;\bX_0,\mu)
\label{evolution}
\end{equation}
and define a norm $\| \bx(t) \|$ to measure the distance of $\bX(t)$
from $\bX_0$. The approach is to find the maximum distance after some
time $t=T$, $\|\bx(T)\|$, over all perturbations which start the same
finite distance 
\beq
\|\bx(0)\|=d
\label{initial_distance}
\eeq
away at time $t=0$ and evolve under
(\ref{evolution}). For the Euclidean norm,
$\|\bx\|_2:=\sqrt{\sum_{n=1}^N x_i^2}$, this can be formulated
particularly easily as maximising the Lagrangian
\begin{equation}
\hspace{-0.5cm}
\La=\La(\bx,\bnu,\lambda;\bX_0,d,T):=\| \bx(T)\|_2^2+ \int^T_0 \, \bnu . 
\biggl( \frac{d \bx}{dt}-\bF \biggr) \, dt
+\lambda(\| \bx(0) \|_2^2-d^2)
\label{Lag}
\end{equation}
with $\bnu(t)$ and $\lambda$ acting as Lagrange multipliers to
impose the dynamical constraint (\ref{evolution}) and initial distance
constraint respectively and `$.$' is the usual scalar product. Maximal values of $\La$ are identified by
vanishing first variations with respect to each of $\bx(t)$, $\bnu(t)$
and $\lambda$ (\,$\bX_0$, $d$ and $T$ are fixed\,). The
first variation of $\La$ with respect to $\bx(t)$ is
\beqa
\hspace{-1cm}\delta \La&:=& \lim_{\eps \rightarrow 0}
\frac{\La(\bx+\eps \delta \bx,\bnu,\lambda; \bX_0,d,T)-\La(\bx,\bnu,\lambda;\bX_0,d,T)}{\epsilon}
\nonumber \\
&=&\left[\,2\bx(T)+\bnu(T) \, \right] \cdot \delta \bx(T)-\int^T_0 \left[ \frac{d\bnu}{dt}
+\bnu \cdot \frac{\partial \bF}{\partial \bx} \right] \cdot \delta \bx \, \, dt
\nonumber \\
&& \hspace{6.5cm}+\left[\,2 \lambda \bx(0)-\bnu(0) \, \right]\cdot \delta \bx(0)
\eeqa
which only vanishes for all allowed variations $\delta \bx(t)$ if 
\beq
\frac{d\bnu}{dt}
+\bnu \cdot \frac{\partial \bF}{\partial \bx} = {\bf 0}  \qquad {\rm over} \, \, t \in (0,T) \label{x(t)}
\eeq
{\it and}
\beq
2 \lambda \bx(0)-\bnu(0) = 2\bx(T)+\bnu(T) = 0. 
\label{x(0)_x(T)}
\eeq 
Stationarity of $\La$ with respect to the Lagrange multipliers
$\bnu$ and $\lambda$ by construction imposes the evolution equation
(\ref{evolution}) and the initial distance constraint
(\ref{initial_distance}) respectively.  Maximising $\La$ is then a
problem of simultaneously satisfying (\ref{evolution}),
(\ref{initial_distance}), (\ref{x(t)}) and (\ref{x(0)_x(T)}). This is
in general a nonlinear system that needs to be solved iteratively.
The solution technique starts with an initial guess $\bx(0)$ which is
integrated forward in time using (\ref{evolution}) to produce
$\bx(T)$. This initializes (via (\ref{x(0)_x(T)})\,) the {\em
  backward} integration of the `dual' or `adjoint' dynamical equation
\begin{equation}
\frac{d \bnu}{dt}=-\bnu.\frac{ \delta \bF}{\delta \bx}
\end{equation}
(with $\delta \bF/\delta \bx$ a matrix and generally dependent on
$\bx(t)$) from $t=T$ back to $t=0$ to generate $\bnu(0)$. At this
point, only two conditions remain to be satisfied.  The Frech\'{e}t
derivative $\delta \La/\delta \bx(0):=2 \lambda \bx(0)-\bnu(0)$ will
{\it not} in general vanish so the strategy is to move $\bx(0)$
(subject to the initial distance constraint) until it does. By
choosing to `ascend' (moving $\bx(0)$ in the direction of $\delta
{\cal L}/\delta \bx(0)$\,) a maximum in $\La$ is sought. The value of
$\lambda$ is simultaneously specified by ensuring that $\| \bx(0)
\|_2=d$ continues to hold during this adjustment in $\bx(0)$.  This
procedure is repeated until $\|\delta \La/\delta \bx(0)\|_2$ is
sufficiently small to indicate a maximum. Theoretically, when $T
\rightarrow \infty$, the global maximum
\begin{equation}
\max_{\bx(0)} \La= \left\{ \begin{array}{cl}
0                      &  \hspace{0.5cm} d < d_c,\\
\|\bX_s-\bX_0\|^2_2    &  \hspace{0.5cm} d=d_c, \\
\|\bX_1-\bX_0\|^2_2    &  \hspace{0.5cm} d > d_c
\end{array}\right.
\end{equation}
where $\bX_1$ is another stable state of the system, $\bX_s$ is a
saddle embedded in the basin boundary with its only unstable manifold
perpendicular to the boundary and in general $\|\bX_s-\bX_0\|_2 \neq
\|\bX_1-\bX_0\|_2 $ (for simplicity $\bX_s$ and $\bX_1$ are assumed
steady).  This is just the statement that depending on whether the
initial perturbation is strictly in the basin of attraction of
$\bX_0$, on the basin boundary or outside the basin (and in the basin
of attraction of $\bX_1$), the endstate is $\bX_0$, $\bX_s$ or $\bX_1$
respectively for large times. The jump in the endstate distance is
then discontinuous once $d$ reaches $d_c$ ({\it a priori} unknown) which
signals that the basin boundary has been reached. The optimal
disturbance $\bx(0)$ at $d=d_c$ has been christened the `minimal seed'
\cite{PK10,PWK12} since it is arbitrarily close to a disturbance which
will trigger the transition from $\bX_0$ to $\bX_1$ for $d=d_c^+$. Put
another way, the minimal seed, renormalised by a factor $1+\epsilon$
where $\epsilon$ is vanishingly small but non-zero, represents the
easiest/most efficient way to leave the basin of attraction of
$\bX_0$. Practically, this behaviour can be mimicked by choosing $T$
large enough that trajectories starting sufficiently close to the
basin boundary but either side of it can be observed to diverge. The
minimal seed is then approximated by the first initial condition as
$d$ is increased which is observed to have escaped the basin of
attraction of $\bX_0$.

%
%
\begin{figure}
\begin{center}
\includegraphics[trim= 5cm 5cm 3cm 2cm,clip=true,width=1.0\columnwidth]{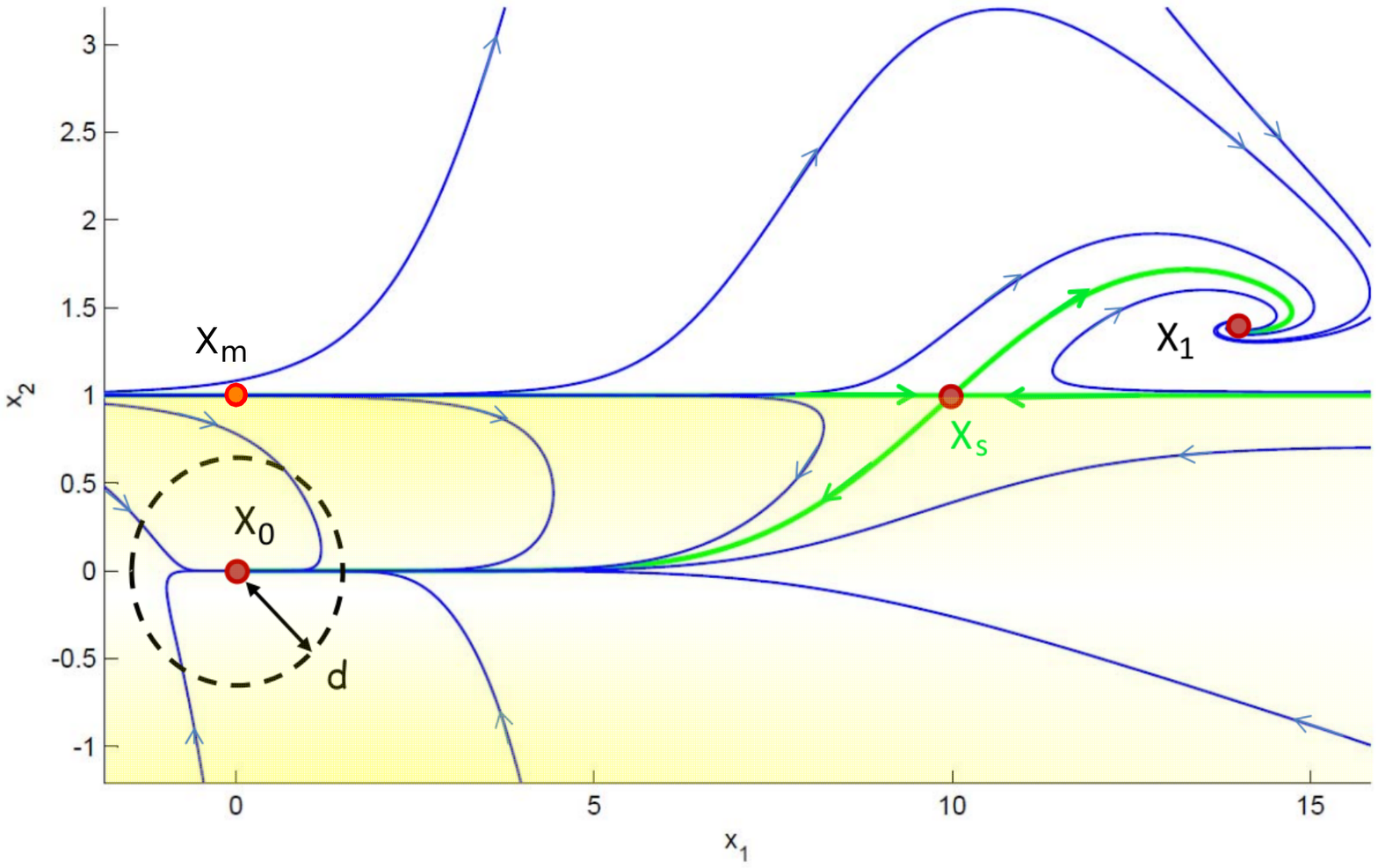}
\end{center}
\caption{Phase portrait for the system defined by (\ref{ODE}). The
  stable and unstable manifolds of $\bX_s$ are drawn in thick green.
  A circle of radius $\|\bx(0)\|_2=d$ about ${\bf X}_0$ is shown as a
  dashed line. As the radius increases, the circle first touches the
  basin boundary at $\bX_m=(0,1)$ - the {\em minimal seed} since
  it is arbitrarily close to a state which will trigger the new
  asymptotic state $\bX_1$ when the system is in state $\bX_0$.}
\label{ODEplot}
\end{figure}

\subsection{A simple example} 

We show how this technique works for the simple
two-dimensional system
\begin{equation}
\frac{d \bX}{dt}=\left[ 
\begin{array}{c} 
-X_1+10X_2\\
X_2(10e^{-X_1^2/100}-X_2)(X_2-1)
\end{array}\right]
\label{ODE}
\end{equation} 
where $\bX=(X_1,X_2)$ which has two local attractors at
$\bX_0=(0,0)$ and $\bX_1=(14.017,1.4017)$. The basin boundary is
simply $X_2=1$ which is the stable manifold of the saddle point at
$\bX_s=(10,1)$: see Figure \ref{ODEplot}.  All initial conditions with
$X_2 <1$ are attracted to $\bX_0$, those with $X_2>1$ to $\bX_1$, and
those with $X_2=1$ to $\bX_s$. Focussing on the `unexcited' state
$\bX_0$, $\bx:=\bX-\bX_0$ and using the Euclidean norm again for
simplicity, it is clear that drawing a circle in the $(x_1,x_2)$ plane
centred on the origin and of increasing radius $d$, the basin boundary
for $\bX_0$ will be first touched at $d=d_c=1$ where $(x_1,x_2)=(0,1)$:
this is the {\em minimal seed} for this system.  As $d$ increases
beyond $d_c$, the basin boundary is increasingly punctured so that
ever more initial conditions are outside of the basin of attraction of
$\bX_0$.

%
%
\begin{figure}
\begin{center}
\includegraphics[trim= 4cm 9.75cm 4cm 9.5cm,clip=true,width=0.45\columnwidth]{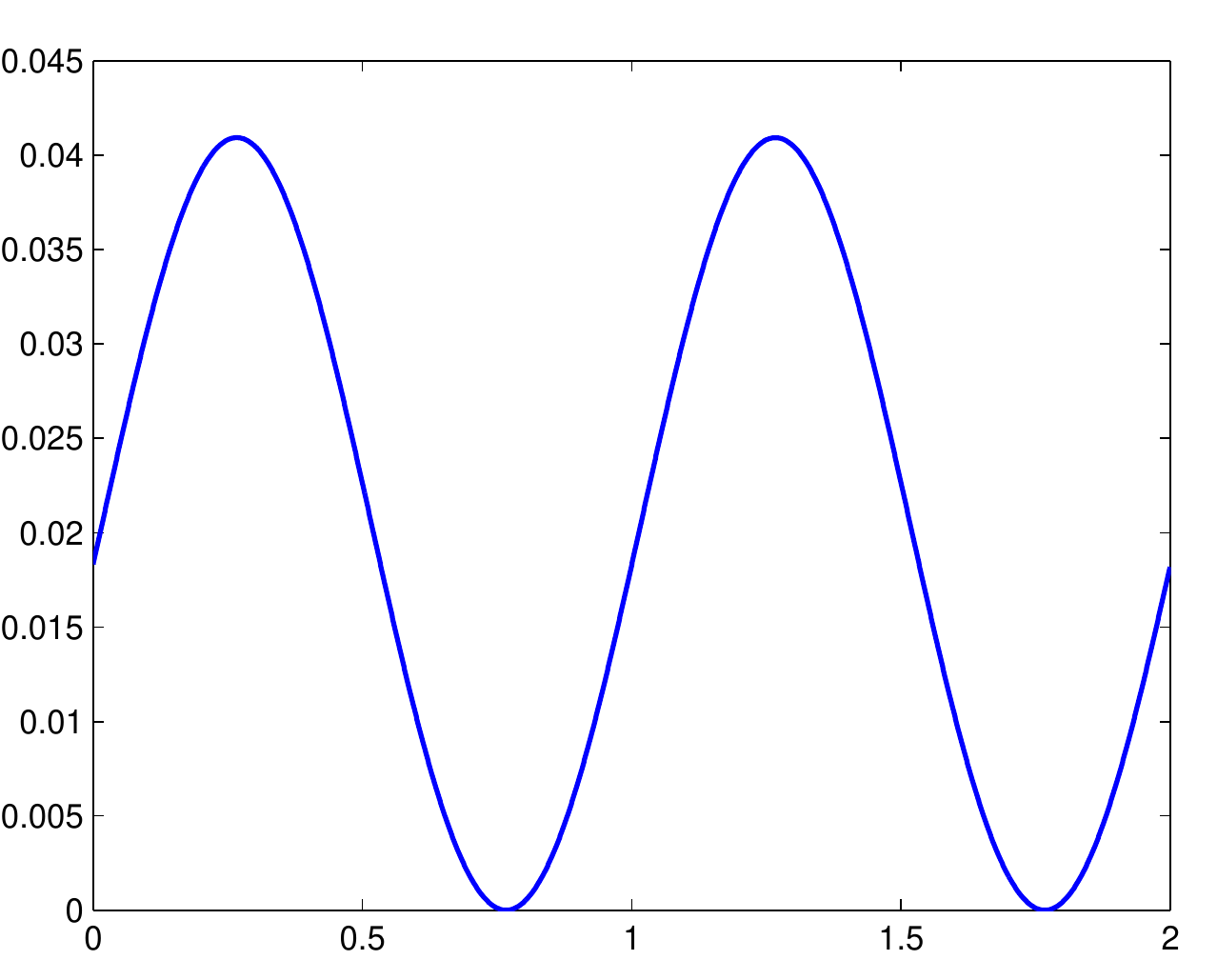}
\includegraphics[trim= 4cm 9.75cm 4cm 9.5cm,clip=true,width=0.45\columnwidth]{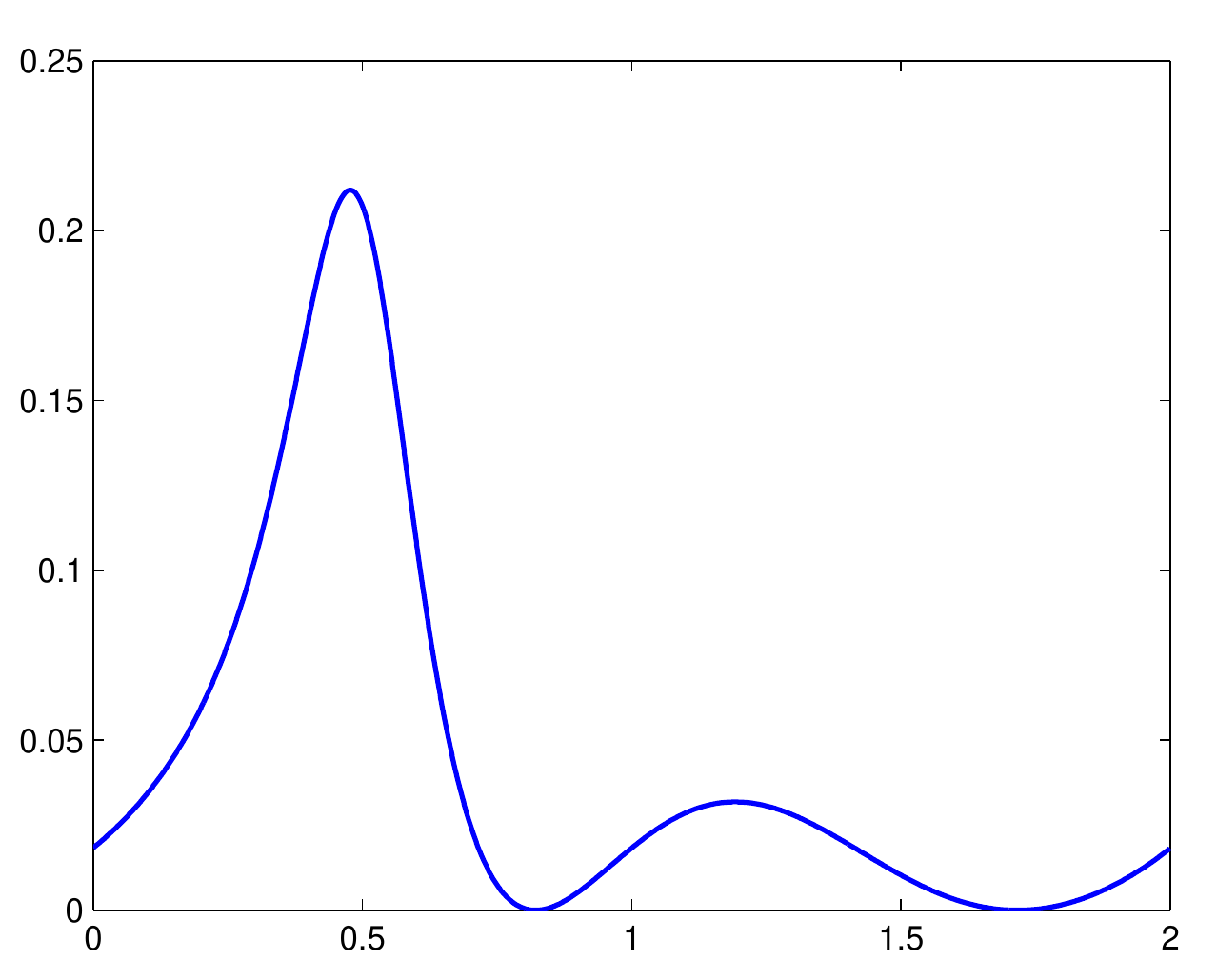}\\
\includegraphics[trim= 4cm 9.75cm 4cm 9.5cm,clip=true,width=0.45\columnwidth]{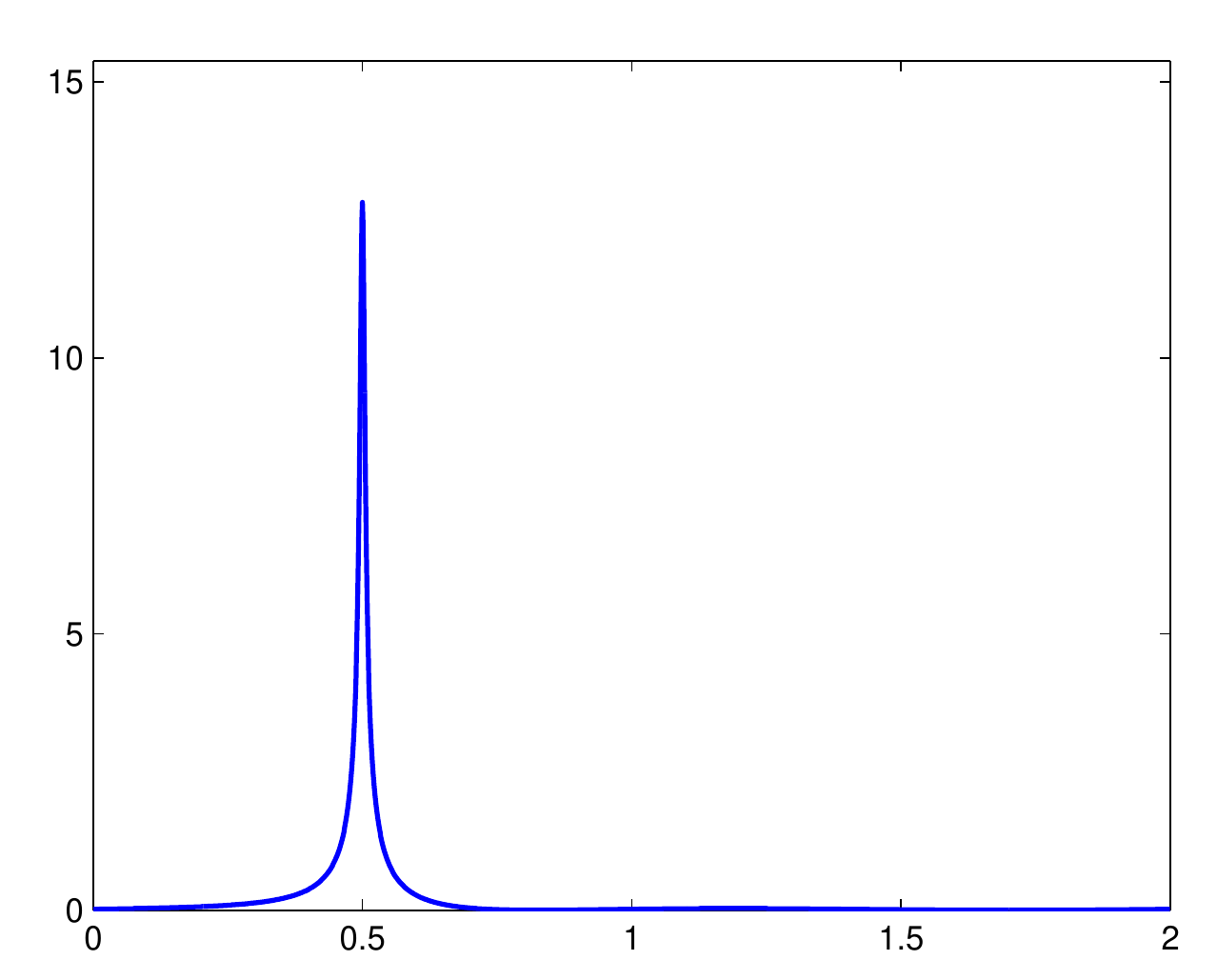}
\includegraphics[trim= 4cm 9.75cm 4cm 9.5cm,clip=true,width=0.45\columnwidth]{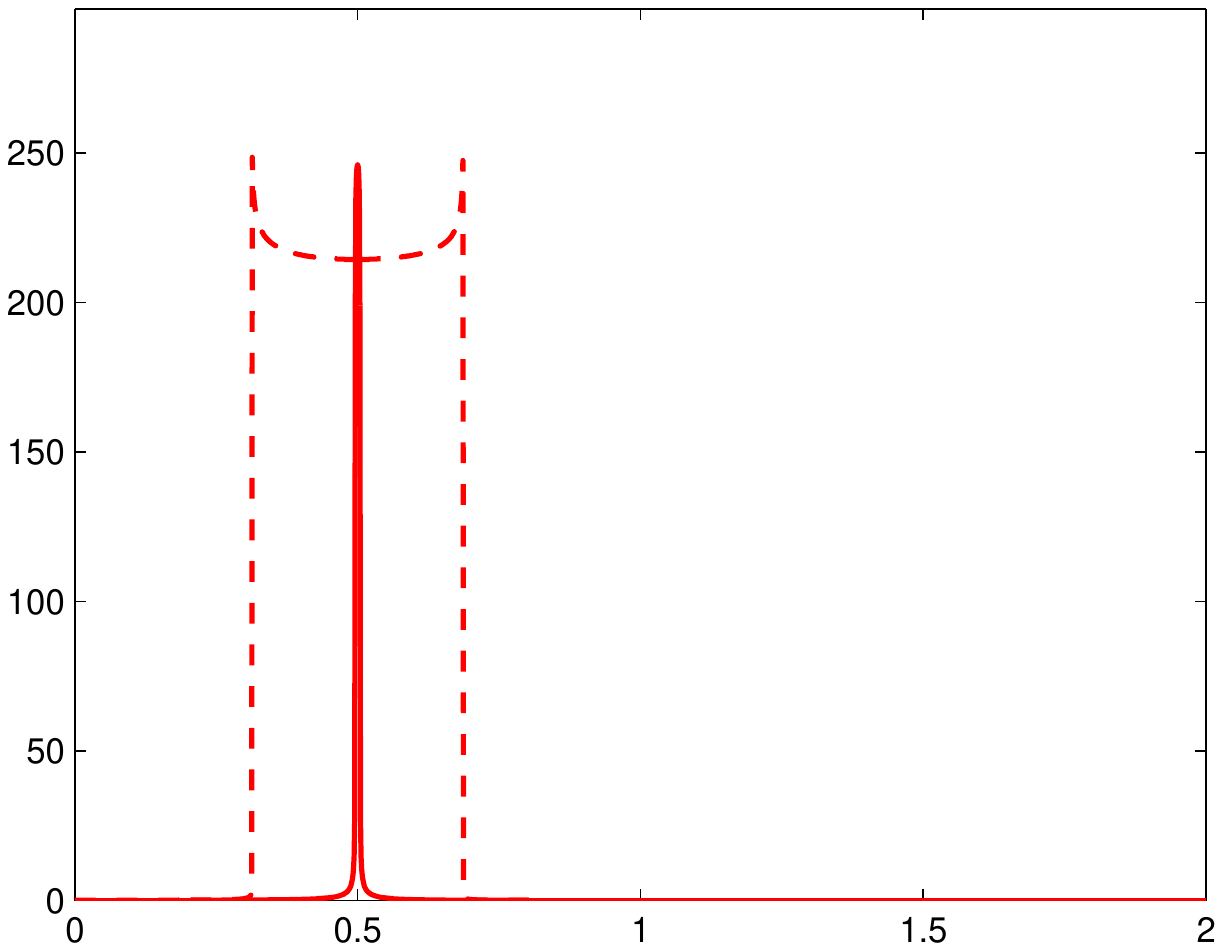}
\end{center}
\caption{Top row: Gain $G:=\|\bx(t)\|_2^2/d^2$ verses $\theta$ for
  $T=2$ and $d=10^{-4}$ (top left) or $0.9$ (top right). Bottom row:
  $\|\bx(T)\|^2_2$ verses $\theta$ for $T=2$ and $d=0.9999$ (bottom
  left) or $1.0001$ (solid line, bottom right) and $1.2$ (dashed line,
  bottom right). Notice the sudden jump in $\|\bx(T)\|_2^2$ when $d$
  increases from $0.9999$ to $1.0001$ signalling that the basin
  boundary of $\bX_0$ has been crossed (\,$\|\bx(T)\|_2^2$
  is simply $100(1-e^{-T})^2+1$ for $d=d_c$\,).}
\label{GvsTheta}
\end{figure}
%
%
\begin{figure}
\begin{center}
\includegraphics[trim= 4cm 16.5cm 4cm 4cm,clip=true,width=15cm, height=9cm]{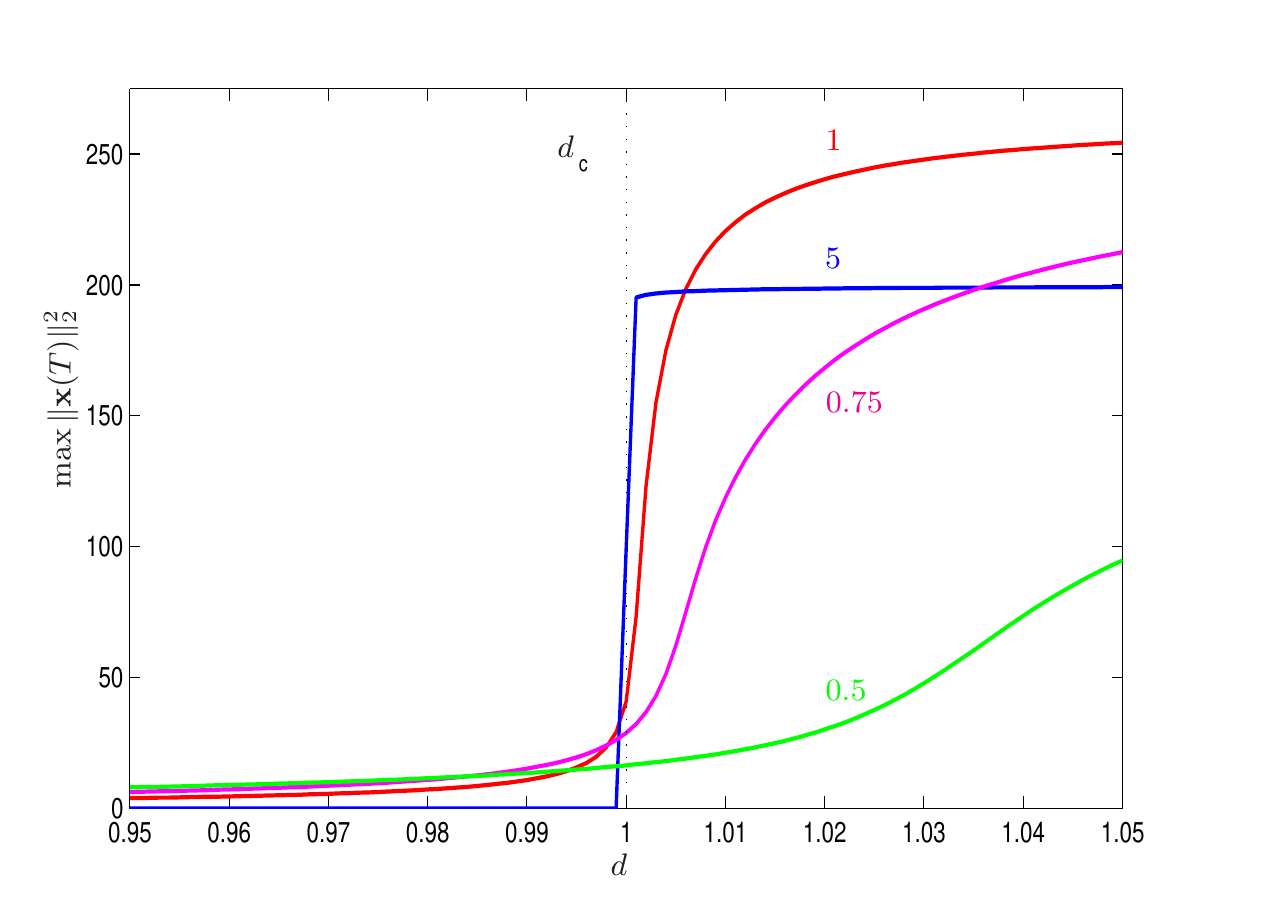}
\end{center}
\caption{As $T$ (labelled on each curve) increases, the signature of
  the transition at $d_c$ becomes increasingly clear. Note only for $T
  \rightarrow \infty$ is the final plateau $\|\bx(T)
  \|_2^2=\|\bX_1-\bX_0\|_2^2\approx 198.5$ as $\bx(t)$ `overshoots':
  see Figure \ref{ODEplot}.  }
\label{GvsE0}
\end{figure}

This observation can be deduced by iteratively solving the
optimisation problem described above through integrating (\ref{ODE})
forwards (since here $\bx=\bX$ as $\bX_0={\bf 0}$) to find $\bx(t)$ and the dual
dynamical equation
\begin{equation}
\frac{d \bnu}{dt}=\left[ 
\begin{array}{c} 
\nu_1+\frac{1}{5}x_1 \nu_2(x_2^2 \!-\! x_2)e^{-x_1^2/100}\\
\! \!-10\nu_1 \!\!+\nu_2(3x_2^2 \!- \! 2x_2)\!-\!10\nu_2(2x_2 \!-\!1)e^{-x_1^2/100}
\end{array}\right]
\end{equation}
for $\bnu(t)$ backwards (note that this backwards-in-time
integration requires knowledge of $\bx(t)$ across $[0,T]$).  Figure 2
shows how $\|\bx(T)\|^2_2$ varies over all permitted initial
conditions parameterised by the angle $\theta \in [0,2)$ where
\begin{equation}
\bx(0)=(\, d\cos \pi \theta, d \sin \pi \theta \,)
\end{equation}
for selected values of $d$ and $T=2$ (note the gain
$G:=\|\bx(T)\|_2^2/d^2$ is plotted for the top row of subplots since
$d$ has such different values). The small value $d=10^{-4}$ is used to
reproduce the linear result obtainable via standard matrix
manipulations (see Appendix A) which possesses the $\bx(0)
\rightarrow -\bx(0)$ ($\theta \rightarrow \theta \pm 1$)
symmetry. As $d$ is increased, this symmetry is quickly destroyed as
nonlinearity becomes important to indicate a unique optimal value of
$\theta$ approaching $0.5$ as $d \rightarrow d_c$. Once $d$ reaches
$d_c$, $\|\bx(T)\|^2_2$ jumps in value as the system explores the
basin boundary and then (for $d>d_c$) the basin of attraction of
$\bX_1$. Three key points highlighted by this simple example are as
follows.
\begin{enumerate}
\item The importance of $T$ in determining $d_c$ to a required
  accuracy. Figure \ref{GvsE0} shows that $d_c$ is increasingly well
  identified as the value of $T$ increases or if $T$ is too small, the
  jump in $\|\bx(T)\|_2$ at $d=d_c$ is smeared out and so difficult to
  locate. On a practical level, $T$ should not be chosen too large as
  the optimisation procedure becomes increasingly sensitive to changes
  in the starting state.
\item The minimal seed is {\em not} related to the linear optimal. For
  `large' $T$ ($\geq 2$), the minimal seed has $\theta^*=0.5$ ($90^o$)
  whereas the linear optimal ($d \rightarrow 0$) has $\theta^* \approx
  0.2667$ ($48.0^o$).  This is not surprising since phase space
  immediately around $\bX_0$ is generally unrelated to the structure
  of the basin boundary a finite distance away. (The one caveat to
  this is if the nonlinear terms had been chosen to be
  energy-preserving - e.g. the model system studied in \cite{DM97}.
  In this case, the dynamics are so tightly constrained in 2
  dimensions that the minimal seed has to be the same as the linear
  optimal: see Appendix B for further discussion.)
\item The Euclidean norm works well as there is little growth for
  trajectories with $x_2<1$, whereas trajectories in $x_2>1$ overshoot
  $\bX_1$. If this were not the case, the functional to optimise - the
  objective functional - could be redesigned to signal the arrival in
  a new basin. Here for example, the value of $x_2(T)$ would be more
  appropriate with the Euclidean norm still used to constrain the
  competitor set of initial perturbations, i.e. the measure used as
  the objective functional and the norm constraining the initial
  condition do not need to be the same.
\end{enumerate}


\section{The Transition to Turbulence problem in Fluid Mechanics}\label{transition}

We now move the discussion to fluid mechanics and a system described
by a PDE.  The breakdown of laminar shear flows has been a central
problem in fluid mechanics since the inception of the subject and
fascinated many generations of scientists (e.g. Rayleigh \cite{R1880},
Kelvin \cite{T1887}, Reynolds \cite{R1883a,R1883b}, Orr \cite{O1907},
Sommerfeld \cite{S1908}, Noether \cite{N1921}, Taylor \cite{T1923},
Heisenberg \cite{H1924}, Landau \cite{L1944}, Hopf \cite{H48} and see
the textbooks \cite{C61,DR81,D02,SH01}).  Beyond a certain value of
the driving rate (measured by a non-dimensional grouping called the
Reynolds number $Re$ and generated by either imposed boundary motion,
flow rate or pressure gradient), unidirectional shear flows (e.g. flow
through a straight pipe) typically exhibit {\em bi}\,stability where a
linearly-stable simple laminar state coexists with a spatially- and
temporally-complicated turbulent state. The so-called `transition'
problem consists in understanding the physical processes by which
ambient noise (present in any real flow) can trigger the observed
transition of the flow from the laminar to the turbulent state. The
fact that the laminar state is linearly stable means that the
transition process is inherently nonlinear and even now still largely
unexplained.  The transition problem is not only a fascinating
mathematical exercise in PDE theory, but the answers are crucial for
informing attempts to inhibit (e.g. in the aircraft industry) or
enhance (e.g. crucial for mixing processes) the phenonemon in
practical applications.

In the last two decades or so, two complementary approaches to the
transition problem have proved popular. The first - variously labelled
`transient growth', `nonmodal stability theory' or `optimal
perturbation theory' - is a {\em linear} theory explaining how, due to
the non-normality of the linearised evolution operator\footnote{An
  operator/matrix $L$ is non-normal if it does not commute with its
  adjoint/transpose $L^\dag$ i.e. $L L^\dag \neq L^\dag L$.} about
the laminar state, infinitesimal disturbances can experience large but
transient magnification of their energy despite the laminar state
being asymptotically stable (all the eigenvalues indicate exponential
decay with time).  This idea has a long history starting with Kelvin
\cite{T1887} and Orr \cite{O1907} but has really only been
systematically explored from the late 1980s onwards: see
\cite{BB88,F88,G91,BF92,TTRD93,RH93}, the reviews \cite{Gr00,R01,S07}
and the books \cite{SH01,TE05}. The theory works well for interpreting
finite time behaviour such as the flow response to noise immediately
downstream of a pipe expansion \cite{CBB10} but actually says nothing
about asymptotically long time behaviour. Initially, it was suggested
that the presence of large transient growth could `elevate'
infinitesimal disturbances into the nonlinear regime where they could
then become sustained (e.g. \cite{BT95,BDT97}). While this picture,
which emphasizes linear effects over nonlinearity, must be generally
correct, no {\em quantitative} predictions of transition amplitudes
can emerge without a fully nonlinear theory \cite{W95,DM97}.

The second approach to the transition problem, on the other hand, is
fully nonlinear.  This views the flow as a huge dynamical system and
the flow state as an evolving trajectory in a phase space populated by
various invariant sets (exact solutions) and their stable and unstable
manifolds \cite{E02,K05,E07,GHC08,CG10,KUV12}. From this perspective,
`transition' occurs when noise or a disturbance simply nudges the flow
out of the basin of attraction of the laminar state. Making this
observation more predictive has, however, been extremely hard due to
the difficulty of mapping out the basin boundary. Some progress has
been made shadowing the basin boundary (or the more general concept of
an `edge', which includes transient turbulence) {\em forward} in time
using an `edge tracking' technique. This finds `edge states' which are
saddles in the full phase space but attractors on the basin boundary
\cite{IT01,SYE06,SEY07} (e.g. $\bX_s$ in Figure
\ref{ODEplot}). However, these are invariably more distant
(energetically) from the laminar state \cite{PWK12} than other parts
of the basin boundary and therefore less obviously relevant for
choosing an initial transition-triggering disturance.  What is really
needed is a technique to track the basin boundary {\em backwards} in
time to reach regions where it is close to the laminar state.

The gap between these two approaches is therefore one of perspective:
the optimal perturbation theory explains how infinitesimal
disturbances can grow temporarily {\em within} the basin boundary
whereas the dynamical systems approach focusses on what exists {\em
  beyond} the basin boundary. The new optimisation technique discussed
in this article naturally bridges this gap by being able to examine
how the optimal perturbation deviates from the linear optimal as the
amplitude of the starting perturbation is increased until the basin
boundary is reached. The original thinking behind the studies
\cite{PK10,PWK12}, however, was a little different being focussed on
tracing the basin boundary or edge `backwards' in time by posing an
optimisation problem.  If the edge state found by edge tracking is
unique then for all starting states {\em on} the basin boundary, the
initial state which subsequently experiences the largest energy growth
over asymptotically large times (so that all trajectories reach the
edge state) {\em will} be the minimal seed \cite{PWK12}. While this
optimisation problem itself is intractable because the basin boundary
is unknown, it does suggest the tractable problem of finding the
largest energy growth over all starting states of a given initial
energy $E_0$. At precisely $E_0=E_c$, where the basin boundary or edge
touches the energy hypersurface at one velocity state, this
optimization problem considers the growth of this state (the minimal
seed) against the energy growth of all the other initial conditions
below the edge. Given that these latter initial conditions lead to
flows that grow initially but ultimately relax back to the basic
state, the minimal seed will remain the optimal initial condition for
the revised optimisation problem for large enough T \cite{PWK12}.  The
simple choice of the kinetic energy of the disturbance as both the
objective functional {\em and} the constraining norm meant that the
vanishing energy limit $E_0 \rightarrow 0$ lead back to the familiar
linear optimal perturbation calculation. It is now clear other choices
could have been made for the objective functional providing it takes
on heightened values for the turbulent state (e.g energy dissipation
\cite{M11,D13} and \cite{F12,F13} for work considering different
norms).

Before giving an example of the optimisation technique at work in fluid
mechanics, we note that there has been many previous efforts to build
upon (linear) optimal perturbation theory to design better ways to
trigger transition. For example, by showing that the linear optimals
can become unstable at sufficient amplitude (e.g. \cite{Z96,RSBH98}),
or by designing a finite amplitude initial disturbance from a very
small set of physically-motivated `basis' states
(e.g. \cite{VC09,DBL10}), or by searching for optimal deformations of
the {\em base flow} so as to create linear instability \cite{BCL03,BB04,GBN04,BB09}.

\subsection{The optimisation technique in pipe flow}

We now show the optimisation technique in action for the problem of
incompressible fluid flow through a cylindrical pipe.  This is a
classical problem in fluid mechanics studied famously by Reynolds
\cite{R1883a,R1883b} during which he first wrote down the
non-dimensional grouping $Re$ now bearing his name.  The flow can
either be driven by imposing a constant pressure drop across the pipe
or by imposing a constant mass flux through the pipe
(e.g. \cite{Mu11}). We choose the former driving here as the
formulation is slightly simpler (the latter is treated in
\cite{PK10,PWK12}): the two cases are equivalent for $L \rightarrow
\infty$ and localised disturbances.  The set up is an incompressible
fluid of constant density $\rho$ and kinematic viscosity $\nu$ flowing
in a circular pipe of radius $s_0$ under the action of a constant
pressure drop imposed across the pipe of
\beq
\Delta p^*=-\frac{4 \rho \nu W}{s_0} L 
\eeq
(where the pipe is $L$ radii long) and the (basic) Hagen-Poiseuille solution to the Navier-Stokes equations is
\beq
\vel^*_{\rm lam}:= W\biggl(1-\frac{s^{*2}}{s_0^2} \biggr) \bz^*, \qquad p^*_{\rm lam}:=-\frac{4 \rho \nu W}{s_0^2} z^*
\eeq
in the usual cylindrical coordinates $(s^*,\phi,z^*)$ aligned with
the pipe axis. Non-dimensionalizing the system using the
Hagen-Poiseuille centreline speed $W$ (e.g. $\vel_{\rm
  lam}:=\vel^*_{\rm lam}/W$) and the pipe radius $s_0$ (so
$s:=s^*/s_0$) gives the Navier-Stokes equations
\beq
\pdd{\vel_{\rm tot}}{t} + \vel_{\rm tot} \cdot\bnabla \vel_{\rm tot} + \bnabla p_{\rm tot} = \frac{1}{Re}\nabla^2 \vel_{\rm tot}
\eeq
where $Re:= s_0 W/\nu$ is the Reynolds number and  $p_{\rm tot}:=p+p_{\rm lam}$ ($p_{\rm lam}:=-4z/Re$) is the total pressure with its deviation, $p$,  from the laminar pressure field periodic 
across the length of the pipe  to maintain the constant pressure drop: i.e. $p(s,\phi,z,t)=p(s,\phi,z+L,t)$. The velocity boundary conditions are
\beq
\vel_{\rm tot}(1,\phi,z,t)={\bf 0} \quad \& \quad 
\vel_{\rm tot}(s,\phi,z,t)=\vel_{\rm tot}(s,\phi,z+L,t),
\label{bcs}
\eeq
the first being no-slip on the pipe wall at $s=1$ and the second 
periodicity across the pipe length. We
consider the energy growth (our `distance' measure) of a
finite-amplitude disturbance
\beq
\vel:=\vel_{\rm tot}-\vel_{\rm lam}
\label{perturbation}
\eeq
to the laminar profile $\vel_{\rm lam}=(1-s^2)\bz$  by defining the Lagrangian
\beqa
\hspace{-1cm}\La &&= \quad {\cal L}(\vel, p, \lambda, \avel, \pi; T, {\color{red}E_0}):=
 \left\langle \frac{1}{2}|\vel(\mathbf{x},T)|^2 \right\rangle
 + \lambda \left\{
      \left\langle \frac{1}{2}|\vel(\mathbf{x},0)|^2 \right\rangle
      - {\color{red} E_0} \right\} \nonumber\\
 && \qquad + \int_0^T
      \left\langle
       \avel(\mathbf{x},t)\cdot
        \left\{
        \pdd{\vel}{t} + (\vel_{\rm lam}\cdot\bnabla) \vel
            + (\vel\cdot\bnabla)\vel_{\rm lam}
            + {\color{red} (\vel\cdot\bnabla)\vel}
					\right. \right. \nonumber \\	
 &&  \qquad \qquad \left. \left. + \bnabla p - \frac{1}{Re}\nabla^2 \vel
        \right\}
      \right\rangle dt + \int_0^T
      \left\langle
       \pi(\mathbf{x},t) \bnabla\cdot\vel
      \right\rangle dt. 
\eeqa
where
\beq
\langle \ldots \rangle = \int_0^L \int_0^{2\pi} \int_0^1 \ldots 
 s \mathrm{d}s \, \mathrm{d}\phi \, \mathrm{d}z
\eeq
and $\lambda$, $\avel$ and $\pi$ are Lagrangian multipliers imposing
the constraints that the initial energy is fixed, that the Navier-Stokes equation 
holds over $t\in [0,T]$ and the flow is incompressible (\,their corresponding
Euler-Lagrange equations are respectively:
\beq
\left\langle \frac{1}{2}|\vel(\mathbf{x},0)|^2 \right\rangle
  = {\color{red} E_0}
\eeq
\beq
\pdd{\vel}{t} + (\vel_{\rm lam}\cdot\bnabla) \vel
              + (\vel\cdot\bnabla)\vel_{\rm lam}
              + {\color{red} (\vel\cdot\bnabla)\vel}
            + \bnabla p - \frac{1}{Re}\nabla^2 \vel
  = {\bf 0},
\label{NS}
\eeq
\beq
\bnabla\cdot\vel = 0\,\,\,).
\eeq
The key additions to the well-known linear calculation 
acknowledging the fact that the disturbance is of finite
amplitude are shown in {\color{red} red}. The linearised problem is
recovered in the limit of $E_0 \rightarrow 0$ whereupon the nonlinear
term $\vel\cdot\bnabla \vel$ becomes vanishingly small relative to the
other (linear) terms. On dropping this nonlinear term, the amplitude
of the disturbance is then arbitrary for the purposes of the
optimisation calculation and it is convenient to reset $E_0$ from
vanishingly small to 1. In this case the maximum of ${\cal L}$ is then precisely
the maximum gain in energy over the period $[0,T]$.

The Euler-Lagrange equation for the pressure $p$ is
\beqa
0 &&= \int_0^T \left\langle \frac{\delta \La}{\delta p} \delta p\right\rangle dt
 = \int_0^T \left\langle (\avel\cdot\bnabla)\delta p \right\rangle dt
 \nonumber\\
 &&= \int_0^T \left\langle \bnabla\cdot(\avel \delta p) \right\rangle dt
   - \int_0^T \left\langle \delta p (\bnabla\cdot\avel) \right\rangle dt.
\eeqa
which vanishes if $\avel$ satisfies natural boundary conditions (which turn out here to be the same as those for $\vel$)
and 
\beq
\bnabla\cdot\avel = 0
\label{kimura:eq:dual_incompressibility}
\eeq
(note $\delta p$ is periodic across the pipe to ensure the pressure drop across the pipe is constant). The variation in $\La$ due to $\vel$ (with $\delta\vel$ satisfying (\ref{bcs})~) is
\beq
\hspace{-2cm} \delta \La=
\int_0^T \left\langle \frac{\delta \La}{\delta \vel} \cdot \delta \vel
              \right\rangle
  = \left\langle \vel(\mathbf{x},T) \cdot \delta\vel(\mathbf{x},T)
        \right\rangle
    + \lambda
        \left\langle \vel(\mathbf{x},0) \cdot \delta\vel(\mathbf{x},0)
        \right\rangle 
\eeq
\beq
\hspace{2cm} + \int_0^T
        \left\langle 
        \avel \cdot 
        \left\{
         \pdd{\delta\vel}{t} + (\vel_{\rm lam}\cdot\bnabla) \delta\vel
            + (\delta\vel\cdot\bnabla)\vel_{\rm lam} \right. \right.
\eeq            
\beq    
\hspace{1.5cm}   \left. \left.
            +{\color{red} \vel.\bnabla \delta \vel}
            +{\color{red} \delta \vel \cdot \bnabla \vel}
            - \frac{1}{Re}\nabla^2 \delta\vel
         \right\}
        \right\rangle dt
            + \int_0^T
        \left\langle \pi \bnabla\cdot\delta\vel \right\rangle dt.
\eeq
The first term in the second line of
the above equation can be reexpressed as
\beqa
\hspace{-2.5cm}
\int_0^T \left\langle \avel \cdot \pdd{\delta\vel}{t}
   \right\rangle dt
&&=
 \int_0^T \left\langle \pdd{ }{t}(\delta\vel\cdot\avel) \right\rangle dt
- \int_0^T \left\langle \delta\vel \cdot \pdd{\avel}{t} \right\rangle dt
\nonumber \\
&&=
 \left\langle \delta\vel(\mathbf{x},T)\cdot\avel(\mathbf{x},T)
   - \delta\vel(\mathbf{x},0)\cdot\avel(\mathbf{x},0) \right\rangle
 - \int_0^T \left\langle \delta\vel \cdot \pdd{\avel}{t} \right\rangle dt,
\eeqa
the second term as
\beqa
\left\langle \avel \cdot \left\{
  (\vel_{\rm lam}\cdot\bnabla)\delta\vel \right\} \right\rangle
&&=
 \left\langle \bnabla\cdot((\avel\cdot\delta\vel)\vel_{\rm lam})
  - \delta\vel \cdot \left\{(\vel_{\rm lam}\cdot\bnabla)\avel \right\}
   \right\rangle \nonumber\\
&&= - \left\langle
    \delta\vel \cdot \left\{(\vel_{\rm lam}\cdot\bnabla)\avel \right\}
     \right\rangle,
\eeqa
the third term as
\beq
\left\langle \avel \cdot \left\{
  (\delta\vel\cdot\bnabla)\vel_{\rm lam} \right\} \right\rangle
=
\left\langle \delta\vel \cdot \left\{
  \avel\cdot (\bnabla\vel_{\rm lam})^{\rm T} \right\} \right\rangle
\quad (=
\left\langle \delta u_i ~
  \nu_j ~ \partial_i u_{{\rm lam},j} \right\rangle).
\eeq
the fourth and fifth as 
\beq
{\color{red} 
\left\langle
\avel \cdot \left(
\delta \vel \cdot \bnabla \vel+\vel \cdot \bnabla \delta \vel
\right) \right\rangle
=\left\langle
\delta \vel \cdot \left(
[\bnabla \vel]^T \cdot \avel-\vel \cdot \bnabla \avel
\right)
\right\rangle
}
\eeq
and the sixth term as
\beq
\left\langle \avel \cdot
 \left(-\frac{1}{Re}\nabla^2\delta\vel \right) \right\rangle
= -\left\langle \frac{1}{Re} \delta\vel \cdot
    \nabla^2 \avel \right\rangle,
\eeq
and finally the last term as
\beqa
\left\langle \pi \bnabla\cdot\delta\vel \right\rangle
 &= \left\langle \bnabla\cdot \pi \delta\vel \right\rangle
   - \left\langle \delta\vel\cdot \bnabla\pi \right\rangle \nonumber\\
 &= - \left\langle \delta\vel\cdot \nabla\pi \right\rangle.
\eeqa
where $\pi$ has to be periodic as 
$\langle \delta \vel \cdot \bz \rangle \neq 0$ (a change in the mass flux is permitted for constant pressure-drop driven flow) for the surface term to drop. 
Combining all these gives
\beqa
\hspace{-1.5cm}
\int_0^T \left\langle \frac{\delta \La}{\delta \vel} \cdot \delta \vel
              \right\rangle
 &&= \left\langle \delta\vel(\mathbf{x},T)
      \cdot \left\{ \vel(\mathbf{x},T) + \avel(\mathbf{x},T)
            \right\} \right\rangle \nonumber\\
 &&~
   + \left\langle \delta\vel(\mathbf{x},0)
      \cdot \left\{ \lambda\vel(\mathbf{x},0) - \avel(\mathbf{x},0)
            \right\} \right\rangle \nonumber\\
 &&~
   + \int_0^T
      \left\langle
       \delta\vel\cdot
        \left\{
         -\pdd{\avel}{t} 
         - ([\vel_{\rm lam}+{\color{red}\vel}]\cdot\bnabla) \avel
            + \avel\cdot(\bnabla 
            [\vel_{\rm lam}+{\color{red}\vel}])^{\rm T}
         \right. \right. \nonumber \\
 &&~ \hspace{6cm} \left. \left.  
          - \bnabla \pi - \frac{1}{Re}\nabla^2 \avel
        \right\}  \right\rangle dt.
\eeqa
For this to vanish for all allowed $\delta \vel(\mathbf{x},T)$,
$\delta \vel(\mathbf{x},0)$ and $\delta \vel(x,t)$ with $t \in (0,T)$  means
\beqa
\hspace{-2cm}
\frac{\delta \La}{\delta \vel(\mathbf{x},T)} = {\bf 0}
 && ~~ \Rightarrow ~~
   \vel(\mathbf{x},T) + \avel(\mathbf{x},T) = {\bf 0},
   \label{kimura:eq:compatibility_condition_T}  \\
\hspace{-2cm}
\frac{\delta \La}{\delta \vel(\mathbf{x},0)} = {\bf 0}
 && ~~ \Rightarrow ~~
   \lambda \vel(\mathbf{x},0) - \avel(\mathbf{x},0) = {\bf 0},
   \label{kimura:eq:compatibility_condition_0}  
\eeqa
\beq
\hspace{-2cm}
\frac{\delta \La}{\delta \vel} = {\bf 0}
 ~~ \Rightarrow ~~
    \pdd{\avel}{t} + 
    (\vel_{\rm lam}+{\color{red} \vel})\cdot\bnabla \avel
   -\avel\cdot(\bnabla [\vel_{\rm lam}+{\color{red} \vel}])^{\rm T}
          + \bnabla \pi + \frac{1}{Re}\nabla^2 \avel = {\bf 0}.
   \label{kimura:eq:dual_NS}
\eeq
This last equation is the {\em dual} (or adjoint) Navier-Stokes
equation for evolving $\avel$ {\em backwards} in time because of the
negative diffusion term.  This dual equation has the same means of
driving - constant pressure drop - as the physical problem, a
situation also true for the constant mass-flux situation \cite{PK10,PWK12}
 
The approach for tackling this optimization problem is iterative as in
the linear situation \cite{LB98,ABH99,CB00,L00,GSH06} (see also the
review \cite{LB14}), the nonlinear calculation of \cite{ZLB04} using
the (parabolic) boundary layer equations and more generally
\cite{Gu00}. It is essentially as outlined in \S 2.
\begin{description}
\item[Step 0.]
Choose an initial condition 
$\vel^{(0)}(\mathbf{x},0)$ such that
\beq
\left\langle \frac{1}{2}
 |\vel^{(0)}(\mathbf{x},0)|^2 \right\rangle = {\color{red} E_0}.
\eeq
\end{description}
The (better) next iterate $\vel^{(n+1)}(\mathbf{x},0)$
is then constructed from $\vel^{(n)}(\mathbf{x},0)$ as follows:
\begin{description}
\item[Step 1.]
Time integrate the Navier-Stokes equation
(\ref{NS}) forward
with incompressibility $\nabla\cdot\vel=0$ and using the boundary conditions (\ref{bcs})
from $t=0$ to $t=T$ with the initial condition
$\vel^{(n)}(\mathbf{x},0)$ to find $\vel^{(n)}(\mathbf{x},T)$.
\bigskip
\item[Step 2.] 
Calculate $\avel^{(n)}(\mathbf{x},T)$ using
  (\ref{kimura:eq:compatibility_condition_T}) which is then used as
  the {\em initial} condition for the dual Navier-Stokes equation
  (\ref{kimura:eq:dual_NS}).
\bigskip
\item[Step 3.]
Backwards time integrate the dual Navier-Stokes equation
(\ref{kimura:eq:dual_NS})
with incompressibility (\ref{kimura:eq:dual_incompressibility})
and boundary conditions (\ref{bcs})
from $t=T$ to $t=0$ with the `initial' condition
$\avel^{(n)}(\mathbf{x},T)$ to find $\avel^{(n)}(\mathbf{x},0)$.
\bigskip
\item[Step 4.] 
Use the fact that 
\beq
\frac{\delta \La}{\delta \vel(\mathbf{x},0)} = \lambda \vel(\mathbf{x},0) - \avel(\mathbf{x},0) 
\label{dL/du0}
\eeq is now computable to move $\vel(\mathbf{x},0)$ towards a maximum
of $\La$.  One approach \cite{PK10,PWK12,RCK12} is to simply move
$\vel^{(n)}(\mathbf{x},0)$ in the direction of maximum ascent of
$\La$, i.e. a correction to $\vel^{(n)}$ is calculated as follows:
\beqa 
\vel^{(n+1)} &&= \vel^{(n)} + \epsilon \left[\frac{\delta
    \La}{\delta\vel(\mathbf{x},0)}\right]^{(n)} \\ &&= \vel^{(n)} +
\epsilon \left( \lambda \vel^{(n)}(\mathbf{x},0) -
\avel^{(n)}(\mathbf{x},0) \right), 
\eeqa 
with $\lambda$ chosen such that 
\beqa 
{\color{red} E_0} &&= \left\langle \half
|\vel^{(n+1)}(\mathbf{x},0)|^2 \right\rangle \\ &&= \left\langle \half
|(1+\epsilon\lambda)\vel^{(n)}(\mathbf{x},0) - \epsilon
\avel^{(n)}(\mathbf{x},0)|^2 \right\rangle.  
\eeqa 
Here $\epsilon$ is a parameter which can be adjusted as the iteration proceeds to improve
convergence (e.g. \cite{PWK12,RCK12}).
\end{description}
%
%
Other strategies have been adopted - e.g. a relaxation approach
\cite{M11,D13} or a conjugate gradient method
\cite{CDRB10,CDRB11,CDRB12,CD13,J11} - but it is presently unclear which,
if any, is superior. 

\begin{figure}
\centerline{\includegraphics[trim= 0cm 0cm 0cm 19cm,clip=true,width=10cm, height=5cm]{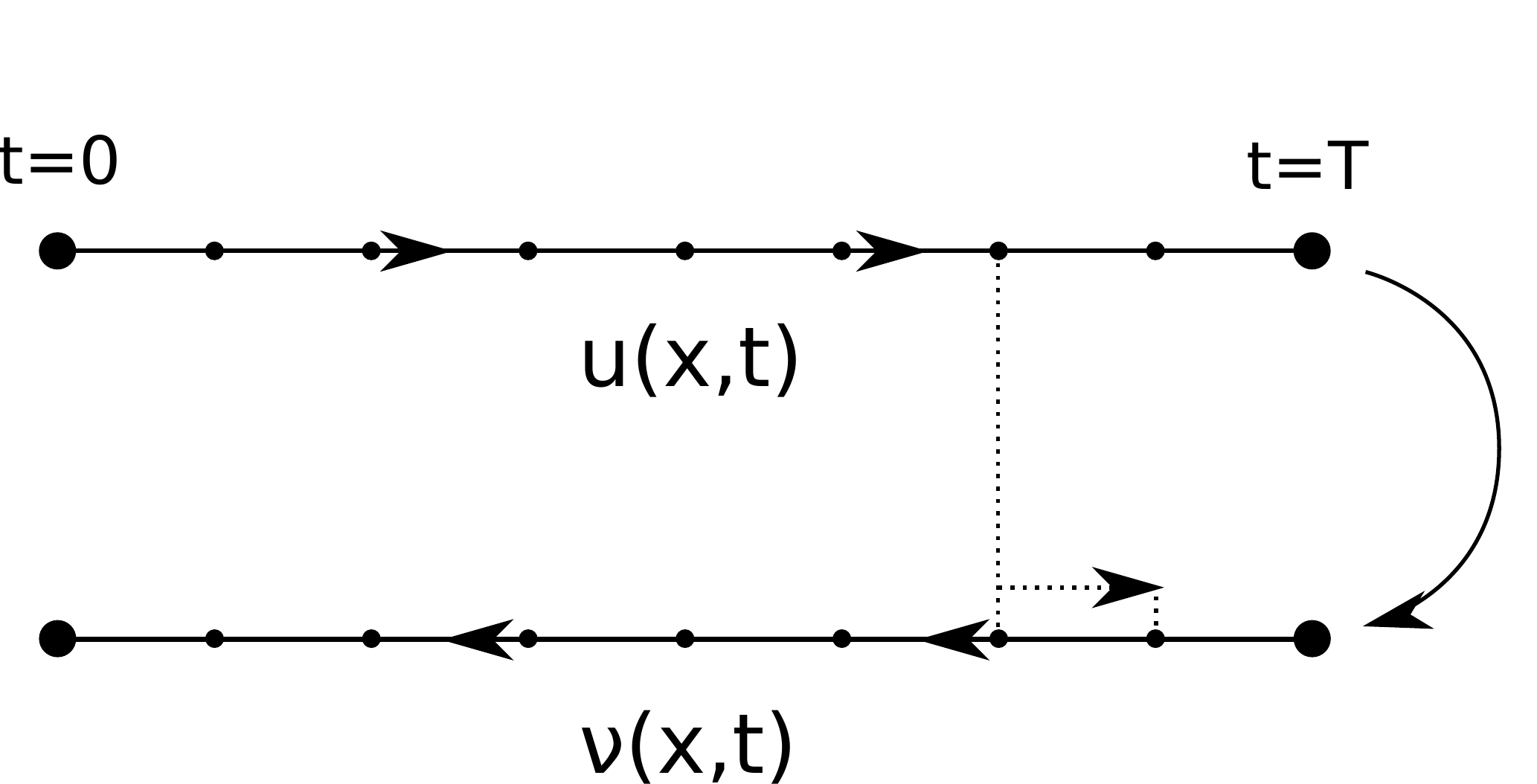}}
\caption{Checkpointing: during the (backward-in-time) calculation of
  $\avel(\mathbf{x},t)$ (indicated by the lower line), the velocity
  $\mathbf{u}(\mathbf{x},t)$ is recalculated in short sections from
  each checkpoint (shown as dots on the forward-in-time calculation)
  where $\mathbf{u}(\mathbf{x},t)$ has been stored during the forward-in-time
  calculation.}
\label{checkpointing}
\end{figure}

%
%
The iterations are repeated until some convergence criterion is
fulfilled.  Some authors \cite{PK10,PWK12,RCK12} have concentrated on
how the residual $\langle (\delta \La/ \delta \vel(\mathbf{x},0) )^2
\rangle$ behaves as a function of the iteration number and others
\cite{CDRB10,CDRB11,CDRB12,CD13} have focussed on the incremental
change in $\La$ between iterations. The former seems more natural
given the latter depends on how large a step is taken in moving
$\vel(\mathbf{x},0)$ but has its issues too (e.g. Figure 9 in
\cite{PWK12}, Figure 9(b) in \cite{RCK12}). More work is needed to
identify a robust convergence criterion.

%
%
There is one further practical issue which needs to be discussed when
the fully nonlinear optimisation problem is considered: the dual
Navier-Stokes equation is linear in $\avel$ but depends on
$\mathbf{u}(\mathbf{x},t)$. This field either needs to be stored in
totality (over the whole volume and time period), which is only
practical for low resolution, short integrations, or must be
recalculated piecemeal during the backward integration stage. This
latter `check-pointing' approach \cite{B98,HWS06} requires that
$\boldsymbol{\bu}$ is stored at regular intermediate points,
e.g. $t=T_i:=iT_{opt}/n$ for $i=1,\ldots,n-1$, during the forward
integration stage. Then to integrate the adjoint equation backward
over the time interval $[T_{i},T_{i+1}]$, $\boldsymbol{\bu}$ is
regenerated starting from the stored value at $t=T_i$ by integrating
the Navier-Stokes equations forward to $T_{i+1}$ again: see
Figure~\ref{checkpointing}. The extent of the check pointing is chosen
such that the storage requirement for each subinterval is
manageable. The extra overhead of this technique is to redo the
forward integration for every backward integration, so approximately a
$50\%$ increase in cpu time, assuming forward and backward
integrations take essentially the same time.  As memory restrictions
may make full storage impossible, this is a small price to pay.

%
%

\begin{figure}
\begin{center}
\includegraphics[trim= 3.5cm 9cm 3.5cm 9cm,clip=true,width=13.5cm, height=11cm]{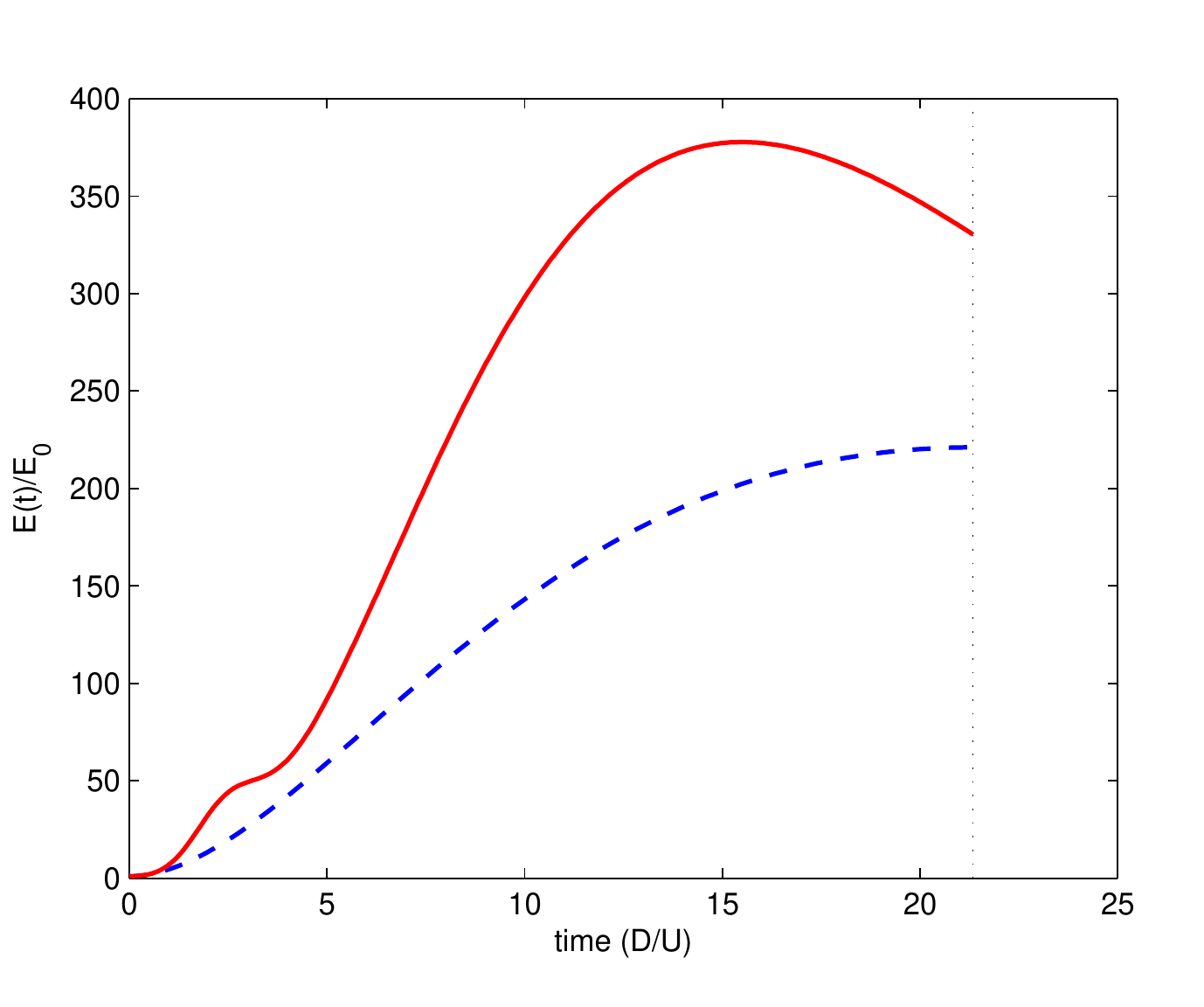}
\end{center}
\caption{Pipe flow: the evolution of the linear and nonlinear energy
  growth optimals at $Re=1750$ in a pipe of length $\pi$ radii
  and constant flux \cite{PK10}. The blue dashed line corresponds to the linear optimal
  for $E_0 \rightarrow 0$ whereas the red solid line is the nonlinear
  optimal for $\approx 1.5E_{3d}$ both calculated using (a short) $T$
  equal to the linear growth optimal time shown as a vertical black
  dotted line (the nonlinear optimal actually produces even more
  growth at a slightly earlier time).  }
\label{Lin_Nonlin}
\end{figure}

\begin{figure}
\begin{center}
\includegraphics[trim= 3cm 9cm 3.5cm 9cm,clip=true,width=13.5cm,height=11cm]{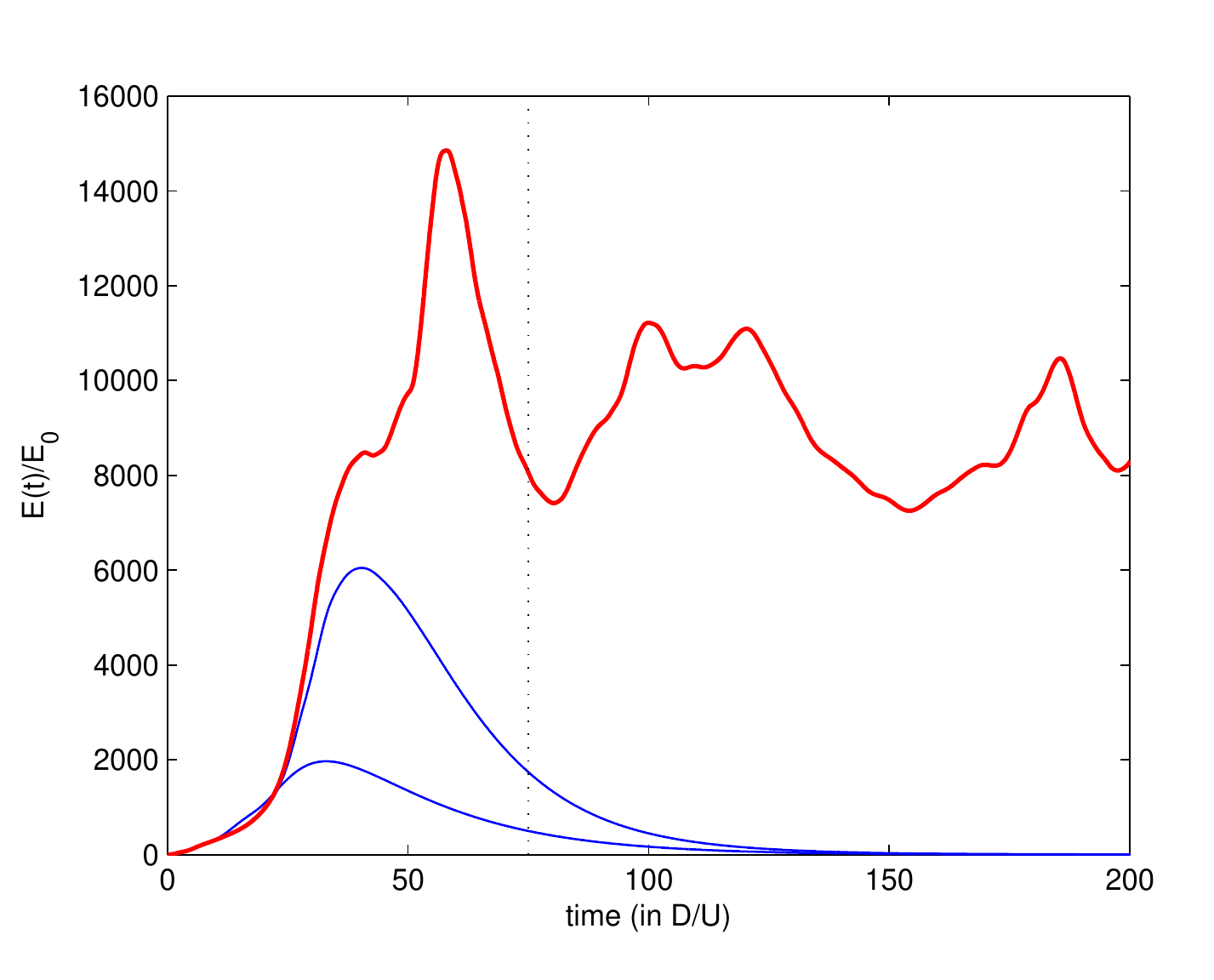}
\end{center}
\caption{Pipe flow: the evolution of the energy growth optimals at
  $Re=2400$ in a constant-flux pipe of length $10$ radii using a `large' time
  $T=75\,D/U$ indicated by a vertical black dotted line (see section 5
  of \cite{PWK12}). The thin blue lines correspond to the NLOPs for
  two $E_0$s just below $E_c$ (within 1\%) with turbulence not
  triggered. The thick red solid line is for $E_0$ just above $E_c$
  (again within 1\%) and turbulence is now clearly triggered (see
  Figure \ref{unpack} for snapshots of this evolution).  }
\label{E_evol}
\end{figure}

\begin{figure}
\begin{center}
\includegraphics[trim= 5.5cm 8.5cm 6cm 8.5cm,clip=true,width=12cm, height=10cm]{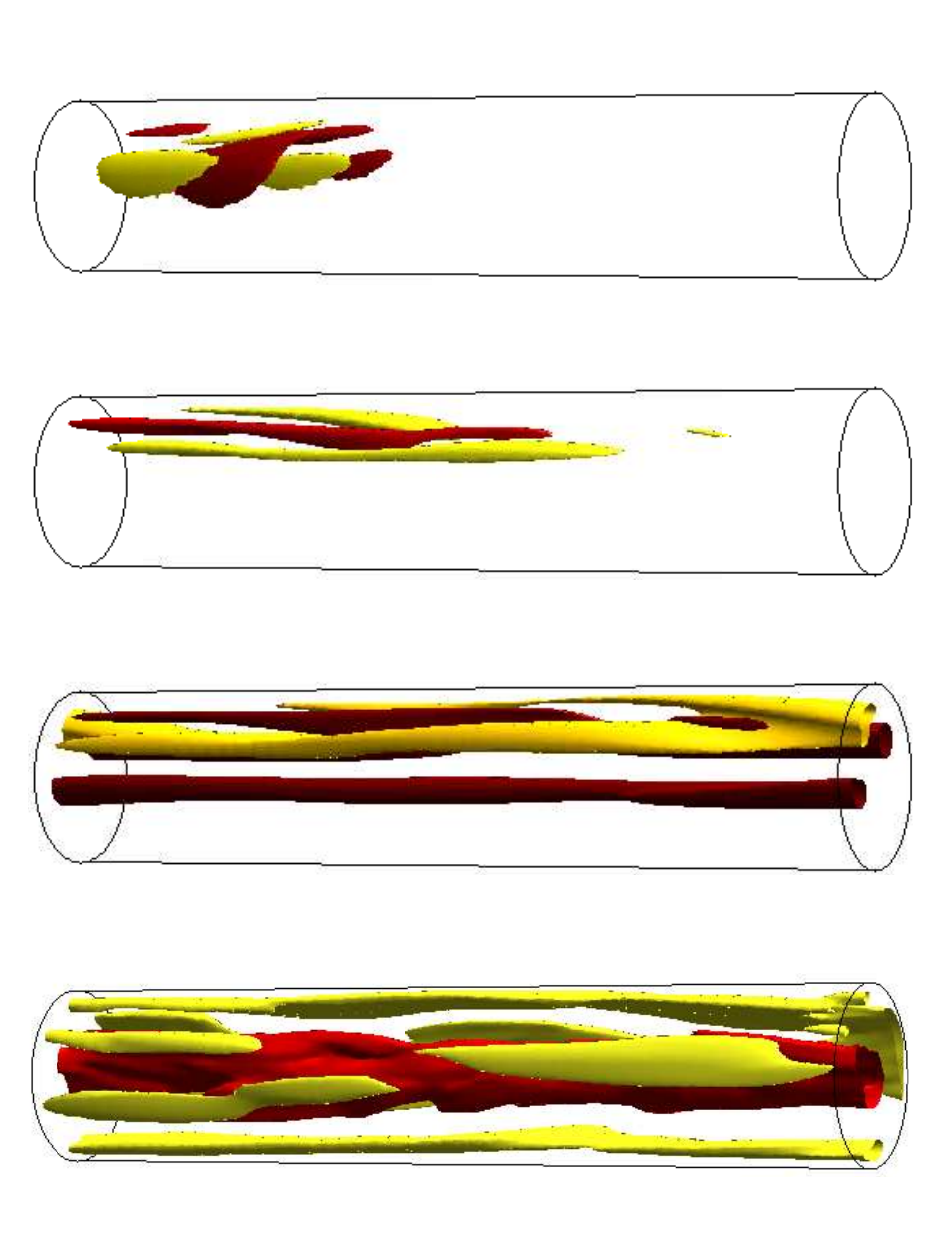}
\end{center}
\caption{Pipe flow: snapshots showing the evolution (time running
  downwards and  fluid flow left to right) of a turbulence-triggering perturbation which
  approximates the minimal seed in a $10$ radii long pipe at $Re=2400$ (constant flux)
  \cite{PWK12} (the corresponding evolution of the total energy is
  shown as the red thick line in Figure \ref{E_evol}). The isocontours
  in each plot correspond to 50\% of the maximum (light/yellow) and
  50\% of the minimum (dark/red) of the streamwise perturbation
  velocity in the pipe at $t=0$ (top: essentially the minimal seed),
  $t=5$ (second down), $t=20$ (third down) and $t=75\, D/U$ (bottom:
  the target time). The perturbation energy is initially localised but
  quickly spreads out to generate streamwise streaks (by $t \approx
  10$) which then break up to generate turbulence.  }
\label{unpack}
\end{figure}

\subsection{Results: Minimal seeds and routes to turbulence}

In fluid mechanics, the kinetic energy of the perturbation $\vel$
(defined in (\ref{perturbation})\,) has invariably been used as the
norm to constrain the competitor initial conditions in the
optimisation problem and typically also used as the objective
functional to be optimised (see \cite{F13} for a counterexample). 
However a different objective functional
can be chosen provided it clearly takes on larger values in the
turbulent state than those reached in the basin of attraction of the
laminar state (e.g. viscous dissipation rate or even the total viscous
dissipation over the period $[0,T]$ - \cite{M11,D13}).

After these choices, the optimisation problem has two operational
inputs - the time horizon $T$ and the initial `distance' $d$ (read
energy $E_0$ for the rest of this section) - beyond a complete
specification of the system parameters (e.g. $Re$ and length of pipe
in pipe flow).

\subsubsection{Large $T$}

The approach for identifying the minimal seed outlined above is to
choose a large fixed value of $T$ and to explore how the global
optimal value behaves as $E_0$ is increased from zero
\cite{PK10,PWK12,RCK12} (complementary work \cite{M11,D13} instead
reduces $E_0$ until no turbulent state is found).  This was first
pursued for the full Navier-Stokes equations in pipe flow \cite{PK10}
where the {\em l}\,inear ($E_0 \rightarrow 0$) energy growth {\em
  o\,}ptimal {\em p\,}erturbation - hereafter referred to as the `LOP'
- is 2-dimensional (streamwise-independent). The optimal for small but
finite $E_0$ is a smooth nonlinear adjustment of the linear result
retaining its 2 dimensionality and with a slightly reduced energy
gain.  Beyond an initial critical energy $E_{3d}$, however, a new
completely different 3-dimensional perturbation was found as the
global optimal and christened the `NLOP' ({\em n}on{\em l}\,inear {\em
  o\,}ptimal {\em p\,}erturbation) \cite{PK10}: see Figure
\ref{Lin_Nonlin}.  This NLOP reflects a clear strategy by the fluid to
spatially localise the starting perturbation so that its peak
amplitude is larger but only over a limited volume to cheat the global
energy constraint. The original pipe geometry used in \cite{PK10} was
very short so the NLOP is only localised in the radial and azimuthal
directions (see Figure 2(a) in \cite{PK10}). Subsequent computations
in much longer pipe geometries have confirmed that the NLOP fully
localises by also localising in the axial (streamwise) direction too
\cite{PWK12, PWK14}: see Figure \ref{unpack}. Physically, the NLOP
evolves by initially `unpacking' (delocalising) under the influence of
the background shear flow and taking advantage of 3 distinct
well-known linear mechanisms for transient growth: the Orr mechanism
which occurs quickly, oblique wave growth which operates over an
intermediate timescale, and lift-up which occurs over a slow timescale
(see \cite{PWK12,D13} and references therein). These mechanisms are
unrelated in the linear problem i.e. initial conditions exploiting
each are distinct from each other so that the mechanisms do not
combine. However they can communicate with the addition of
nonlinearity which allows each mechanism to `pump prime' the next
generating a much higher overall growth \cite{PWK12,D13}. Figure
\ref{Lin_Nonlin} shows a period of growth in the NLOP curve (the
oblique wave growth stage) which terminates at $t \approx 2.5$ before
further growth (lift up) occurs until $t \approx 16$ (the Orr
mechanism occurs over a much faster timescale - $O(0.1)$ see Figure 1
of \cite{PWK12} - and is therefore hidden in this plot). Appendix B
discusses a very simple model to illustrate this phenomenon (compare
Figures \ref{Lin_Nonlin} and \ref{Gvst}).

As $E_0$ is increased further, this NLOP remains the global optimal
until another critical energy $E_{fail}$ is reached beyond which the
optimisation procedure fails to converge.  Pringle et al. \cite{PWK12}
interpret (their conjecture 1) that this corresponds to the first
energy at which an initial condition can reach the turbulent state
since then the extreme sensitivity of the final-state energy at $T$ to
changes in the initial condition, due to exponential divergence of
adjacent states will effectively mean non-smoothness and prevent
convergence. That the turbulent state has been reached at $E_{fail}$
is easily confirmed by examining the endstates $\vel(\bx,T)$ reached
as part of the iteration algorithm so that $E_{fail}$ is at least an
upper bound on the true critical energy $E_c$, which is the lowest
energy at which turbulence can be triggered.  Arguing that in fact
$E_{fail}=E_c$ requires a belief or hope that the search algorithm
will find any state on the (initial) energy hypersurface which can
reach the turbulent state if it exists. This is never likely to be
proved but can at least be made plausible by rerunning the algorithm
with a variety of initial starting states (e.g. see Figure 14 in
\cite{PWK12}).

Pringle et al. \cite{PWK12} also noticed (their conjecture 2) that the
NLOP converged to the minimal seed as $E_0 \rightarrow E_c^{-}$ for
their choice of the energy functional. It now seems clear that there
is nothing special about using the perturbation energy as an objective
functional merely that it is one of a class of functionals which take
on heightened values for turbulent flows. Providing such a functional
is chosen, the best way to maximise it should be to get as close to
the basin boundary as possible while remaining on the $E_0$
hypersurface since all other states will be sucked into the laminar
state if $T$ is chosen larger than the typical transient growth
time. This is certainly borne out by comparing estimates for $E_c$ and
the corresponding minimal seed found in plane Couette flow using the
total viscous dissipation over $[0,T]$ \cite{M11} or the final energy
growth \cite{RCK12} as objective functionals. One of the motivations
for choosing the total dissipation over the time period in
\cite{M11,D13} was to capture the fastest transition. Finding the
critical energy, however, is a problem in finding just one state which
triggers turbulence so, conceptually, the final viscous dissipation
rate could have been used just as well (although practically, time
averaging the dissipation rate helped smoothen the effect of dealing
with turbulent endstates in their optimisation algorithm).

\subsubsection{Optimising over $T$}

Given the need to choose $T$ `large enough', one natural idea has been
to `optimise it out of the problem' by asking the question `for a
given initial perturbation of energy $E_0$ what is the largest growth
over {\em any} $T$?' Algorithmically, this requires only a small
change but can dramatically alter the results for $E_0 < E_c$. Work by
\cite{RCK12} in plane Couette flow has shown how a LOP achieving large
growth at small times can `mask' the emergence of a NLOP (which has a
longer growth time) as $E_0$ increases (see their Figure 2). When
$E_c$ is reached, however, there is the same leap in optimal value
which is now also accompanied with a sudden leap in the optimal time
$T$ (e.g. see Figures 3 and 10 in \cite{RCK12}). Operationally, this
means that the minimal seed (and $E_c$) can still be found albeit only
when the algorithm fails to converge rather than as a smooth
convergence procedure for as $E_0$ approaches $E_c$ from below.
 
\subsubsection{Small $T$}

Cherubini and coworkers \cite{CDRB10,CDRB11,CDRB12} have computed the
nonlinear optimal energy growth disturbances for small $T$ in boundary
layer flow. This is a particularly challenging system since the flow
is spatially developing and open (flow leaving the computational
domain is not recycled using periodic boundary conditions) making
longer time runs very expensive. Rather than isolating minimal
seeds\footnote{Note Cherubini and coworkers
  \cite{CDRB10,CDRB11,CDRB12} use the term `minimal seed' differently
  to this article: they use it to describe the minimal spatial
  structure found in their NLOPs whereas here it is the minimal energy
  state, when infinitesimally disturbed, to trigger turbulence.},
their focus has been to establish that nonlinear mechanisms can
enhance transition by uncovering energies $E_0$ where the NLOP
subsequently triggers turbulence (when integrated beyond $T$) whereas
the linear optimal, scaled up in energy, does not. By analysing the
transition path of their NLOP in detail, they find a number of
mechanisms believed generic to boundary layer transition
\cite{CDRB11,CDRB12}. Some of these appear to carry over to plane
Couette flow \cite{CD13} when transition is studied again using
short-time optimisation analysis. Here, for small $T$, at least two
scenarios are found: a `highly dissipative' bursting path and
short-path depending on how close the perturbation is initially to the
basin boundary.

\subsection{Summary} \label{summary}

The key points to note in applying this variational approach in fluid
mechanics are as follows.

\begin{enumerate}
\item The choice of objective functional is not important provided it
  takes on heightened values for turbulent flows compared to the
  laminar state. The algorithm works best when there is a good
  separation between functional values attained within the basin of
  attraction and the turbulent state. This is typically the case for
  final time values of the perturbation kinetic energy or viscous
  dissipation rate when the flow domain is large enough, and $Re$ is
  not too close to the critical value at which the turbulent state
  first appears \cite{PWK12}.
\item When seeking the transition threshold $E_c$, $T$ needs to be
  large enough to avoid two problems: a) possible transients with
  large growth caused by some parts of the basin boundary attaining
  large functional values, and b) to allow sufficient times for states
  to reach the turbulent attractor \cite{PWK12,RCK12}.
\item The minimal seed can either be identified smoothly as a
  converged NLOP for $E_0 \rightarrow E_c^{-}$ at fixed but large $T$
  \cite{PWK12} or as the first initial condition encountered by the
  algorithm which triggers turbulence as $E_0$ is increased if $T$ is
  being optimised over \cite{RCK12}.
\end{enumerate}

Finally, we end this section with a note of caution. The approach
relies on the correct identification of the global optimal to a
(non-convex) nonlinear optimisation problem. There is unlikely to ever
been a formal criterion to verify that this has been accomplished (in
contrast to the linearised problem) so care must be taken to collect
supporting evidence for this claim if possible. So far, the results
obtained have been checked for consistency in a number of different
ways:

\begin{enumerate}
\item Robustness: it is clearly good practice to check global
  optimality by seeding the variational algorithm with a suite of very
  different initial states to see if the same optimal emerges each
  time (e.g. Figures 14 and 15 of \cite{PWK12}) or, if not, that the
  new optimal is just a local optimal.
\item Physical plausibility: it is reassuring that the NLOPs found are
  localised and appear to `join up' the various hitherto-unconnected
  linear energy-growth mechanisms during their evolution
  \cite{PWK12,CDRB11,D13}.
\item Using different objective functionals: the cross-checking of
  results between two groups \cite{M11,RCK12} using different
  algorithms and objective functionals has been crucial in
  establishing the feasibility of this approach to realistics problems
  in fluid mechanics.
\item Of all the states on the basin boundary, the minimal seed should
  experience the largest increase in `distance' (read perturbation
  energy) as it evolves up to the edge state and states `close' to the
  minimal seed should evolve to pass close to the edge state before
  either relaminarising or trigger turbulence (e.g. Figure 17 of
  \cite{PWK12}).
\end{enumerate}

\section{Applications}

\subsection{Climate modelling and Weather Forecasting}\label{ocean}

In climate modelling and weather forecasting, the sensitivity of
predictive models to uncertainty in their initial conditions (the
current best guess of the model's state) is of central
importance. Lorenz \cite{L65} was the first to introduce the concept
of singular vectors as a way to analyse how this uncertainty (presumed
small) may grow with time. However, it wasn't until the late 1980s
when the idea really took hold following the work of Farrell
\cite{F89,F90} in atmospherics and the accompanying realisation in
general shear flows of the phenomenon of (linear) transient growth or
(linear) `optimal perturbations'
\cite{BB88,F88,G91,BF92,TTRD93,RH93}. This quickly led to the
implementation of singular vectors to generate ensemble forecasts
(e.g. at the European Centre for Medium-Range forecasting (ECMWF)
\cite{P93,BP95}).

The initial uncertainties, however, needn't be small and their true
behaviour with time may differ significantly from that given by the
evolution operator linearised around the base solution. As a result,
there have been attempts to include some nonlinearity into the problem
by an iterative approach \cite{OB95,B96} as well as a proposal for a
fully nonlinear approach by Mu and coworkers:
\cite{Mu00,MDW03,MZ06,MJ08a,MJ08} and more recently
\cite{DZ13,ZMD13,JML13}.  \cite{MDW03} in particular introduced the
concept of a `conditional nonlinear optimal perturbation' (CNOP),
which is exactly the NLOP discussed above, to study the predictability
of a very simple coupled atmosphere-ocean model. As mentioned in the
Introduction, the idea to study the transitions between different
stable states was also suggested by Mu and co-workers in the context
of ocean modelling where multiple equilibria are a generic feature of
general circulation models. To illustrate this idea, \cite{MSD04}
studied a very simple box model of the thermohaline circulation
consisting of 2 ODEs using CNOPs. Subsequent work has tried to scale
up the calculation of CNOPs to realistic PDE models with \cite{MZ06}
computing CNOPs for a 2D quasigeostrophic model with 512 grid points
and \cite{TD08} treating a 2D barotropic double-gyre ocean flow model
with 4800 degrees of freedom. The latter study, however, concluded
that finding basin boundaries was just too computationally expensive
to attempt. Given the recent successes in the transition problem, this
conclusion probably deserves to be revisited.

Before leaving this section, the work of Toth and Kalnay
\cite{TK93,TK97,K03} on the so called `breeding' method deserves
mention. This consists of adding a small arbitrary perturbation to the
full forecasting model, allowing this to evolve and then rescaling it
after a given time to reseed the next forecast period. This procedure,
which breeds `bred vectors', is less optimal in identifying optimal
perturbations for a given period (in fact perturbations which emerge have
{\em just} grown the fastest rather than {\em will} grow the fastest) 
but is readily generalised to incorporate the finite-amplitude
nature of the uncertainties since the full forecasting model is
used. This method is used to generate ensemble forecasts at the
National Centers for Environmental Prediction (NCEP) (formerly the US
National Meteorological Center).

\subsection{Thermoacoustics}\label{thermo}

The idea of using an optimisation approach to locate a basin boundary
was also introduced to the field of thermoacoustics at the same time
as in the transition to turbulence community. \cite{J11} treated a
simple model of thermoacoustic system - a horizontal Rijke tube -
where the laminar state is a fixed point, the edge state is an
unstable periodic orbit and the `turbulent' state is a stable periodic
orbit (respectively the `lower' and `upper' branches which emerge from
a saddle node bifurcation). In contrast with the strategy outlined
above, \cite{J11} found the minimal seed by looking for the minimal
energy state to reach the (unique) edge state at intermediate times
rather than the minimal energy state that reaches the stable periodic
orbit at long times.  However, the result is the same and was
confirmed in \cite{J11a} using the latter strategy. As in the
transition problem (with now the nonlinear terms not
energy-preserving), the minimal seed is found to be completely
different from the linear optimal. The Rijke tube system is
sufficiently simple (a couple of time-dependent 1-space dimension
PDEs) to perform an exhaustive survey of transient growth
possibilities over amplitude and event horizon $T$ \cite{J11a} as well
as including noise \cite{WJ11}. It is also realistic enough to achieve
some correspondence with experiments \cite{JS13}.

\subsection{Control}

Work in controlling fluid flows has long flirted with fully nonlinear
methods (e.g. \cite{J95,J97,Gu00,BMT01,C02,P02}) but they remain
currently very expensive and probably still impractical
\cite{KB07}. The recent success in identifying NLOPs and
minimal seeds, however, has lead to some new activity in this
direction. \cite{PE13} has recently used the adjoint-based
optimisation procedure discussed here for the full Navier-Stokes
equations to control the 2D boundary layer dynamics over a bump by
blowing and sucking appropriately through the boundaries. This work
has been subsequently extended to 3D \cite{CRD13} where initial
conditions corresponding to the LOP and NLOP have been treated.

In a slightly different vein, a more {\em nonlinearly stable} plane
Couette flow has been designed by imposing spanwise oscillations on
the usual streamwise boundary shearing \cite{RCK14}. This work builds
upon the fact that if the critical energy of the minimal seed can be
found then new boundary conditions can be designed to increase this
energy thereby improving the stability of the base state. While the
choice of imposing spanwise oscillations was motivated by a large body
of experimental and theoretical work (see \cite{RCK14} for
references), it also meant that the base state was no longer steady
but time-periodic. This has implications for the optimisation
procedure which now not only has to search for the optimal initial
perturbation but also the exact time (or phase) during the base flow
period when it should be introduced.

\subsection{Magnetic field generation \& Mixing}

\cite{W12} treats the kinematic dynamo problem looking for the
velocity field of an electrically-conducting fluid which produces the
greatest growth of magnetic field at the end of a time interval
$T$. The set of competitor fields is constrained either by the
total energy or the $L_2$ norm of their strain rate (or equivalently
the viscous dissipation rate) only. By using the optimisation procedure
described here, a lower bound on the magnetic Reynolds number is
identified for a dynamo which is only a $1/5$th of that possible
within the well-studied ABC-class of flows \cite{C70,A11}.

Recently \cite{F14} has considered optimal mixing in 2D
channel (plane Poiseuille) flow using a nonlinear optimisation
approach. The mixing of a passive scalar, initially arranged in two
layers, is considered in a parameter regime where the flow is linearly
stable. Nonlinear-adjoint looping is used to identify optimal
perturbations which lead to maximal mixing in some sense and the
classical Taylor dispersion mechanism (where shear enhances
dispersion) is found to emerge naturally from the calculations.

\vspace{1cm} 
Finally, we close this section by noting that the
optimisation procedure discussed in this article is closely related to
the well known data assimilation procedure wherein a dynamical model
of, and incomplete observations about, a real time-evolving system are
used to constrain the initial state of the model such that the `best'
solution over a given time horizon can be sought. This is achieved by
minimising an appropriate cost functional which penalises the
deviations of the predicted solution away from known observations and
possible uncertainties in the dynamical model.  In contrast, the
optimisation approach discussed here assumes a perfect dynamical model
and seeks to maximise an objective functional over all initial
conditions of a certain size in the absence of constraining
observations. Data assimilation is used extensively in many areas of
the geosciences (such as weather forecasting e.g \cite{D91,K03},
oceanography \cite{B92} and more recently modelling the Earth's dynamo
\cite{F10}\,).

%
\section{Final summary and future directions}
%

This article has been a simple introduction to an optimisation
technique which offers a new way to probe the basin boundary of a
state in a dynamical system. Although the discussion has concentrated
on this well-defined situation for clarity of exposition, the
technique can also usefully be employed for systems with just one
global attractor and at least one long-lived but ultimately-repelling
state (sheared fluid flows with enforced short wavelength dynamics are
prime examples of this situation since then the turbulence seems only
transient e.g. \cite{FE04,S10}). Then the global attractor does not
have a basin boundary but there is instead an `edge' or manifold in
phase space which divides initial conditions which immediately
converge to the global attractor and those that first visit one of the
repellors. The same game can then be played providing a functional can
be identified which is clearly maximised in the target repellor for a
time long enough for the repellor to be reached yet shorter than the
mean lifetime of that repellor. This is in fact probably the situation
in all the fluid flow calculations done so far (e.g. the turbulence in
the 5 diameter long pipe in \cite{PWK12} is actually only transient at
$Re=2400$ but the mean lifetime is so large $\gg 100 D/U$ that it
mimicks an attractor on the timescales of the calculations).

The optimisation technique involves maximising a functional which
takes on much larger values in the target state than in the basin of
attraction of the starting state subject to an initial amplitude
constraint and other constraints which include that the governing
equations are satisfied - see the summary in \S \ref{summary}. The
iterative approach to solving this variational problem is not new (e.g
\cite{Gu00}) nor is the realisation that there should be a sudden jump
in the optimal value as the initial amplitude increases to penetrate
the basin boundary particularly profound. What is noteworthy, however,
is that the technique appears to work for large degree-of-freedom
discretizations of PDE systems approaching practical application.

For the particular application studied here - transition to turbulence
in shear flows - the optimisation approach provides a pleasing
theoretical bridge between the two different theoretical perspectives
of (linear) optimal perturbation theory and the (nonlinear) dynamical
system approach. It is now much clearer how nonlinearity interacts
with the linear transient growth mechanisms to achieve transition at
least close to the amplitude threshold.  Furthermore, the first steps have been
taken to actively use the technique to {\it design} more stable states
by adjusting their driving slightly \cite{RCK14}.

There are practical issues, however, needing development. The
optimisation techniques being used to maximise the objective function
given gradient information with respect to the initial conditions are
typically simple-minded with no attempt made so far to tailor
them to the system being treated. There is also no consensus as yet on
what convergence criteria should be used or {\em a posteriori} checks
to confirm that the global maximum has been found. The last issue is
particularly important of course and an agreed level of care is
needed.

\subsection{Future directions}

The optimisation approach is incredibly flexible and there is no
reason why other information cannot be sought from a dynamical
system. For example, \cite{PWK12} talks about identifying the peak
instantaneous pressure in a transitional fluid flow, which is of key
concern in certain applications (e.g. pipeline structural integrity).
Also, so far, only the basin boundaries of steady and time-periodic
states have been probed using this technique. Demonstrating
feasibility for a state with more exotic time-dependence remains to be
done. 

In terms of making greater connection with experiments, the competitor
set of initial states can also be restricted to acknowledge the fact
that only a reduced subset of all initial conditions are achievable in
the laboratory. This can be accomplished simply by projecting the
gradient vector of the objective functional onto a reduced set of
realisable initial conditions and looking for the minimal seed in this
subclass of disturbances.  Looking to make contact with real systems also
highlights a considerable simplification implicit in the discussion so
far: the assumption has been that the system is disturbed {\em once}
and then evolves perfectly. In practice, disturbances to real systems
are not isolated or indeed equally likely. Developing the optimisation
technique to encompass these realities will obviously be important. Some
tentative steps have already been made by considering multiple discrete
disturbances \cite{LK13} and examining how adding noise to a system can
still pick out the minimal seed route to `transition' \cite{WJ11}.

\vspace{1cm}  
Hopefully, the power of the optimisation technique discussed here
should be clear now. Ever increasing computer power is making direct
simulations of systems more common with the concomitant need to
process and interpret this data crucial.  The optimisation
technique discussed here has a huge potential to help with this
and should surely become a standard theoretical tool in the near future.


\vspace{1cm}
\noindent
{\it Acknowledgements}\\ 

RRK would like to thank Colm Caulfield and Sam Rabin for helping to
build his understanding of the subject matter through their joint
work.  Many thanks are also due to Matthew Chantry, Matthew Juniper
and Daniel Lecoanet for commenting on an earlier draft of this
article, and to Stefania Cherubini and Dan Henningson for suggesting appropriate 
control theory references. CCTP acknowledges the support of EPSRC during his PhD.
\vspace{2cm}

\setcounter{section}{1}
\section*{Appendix A: The 2D linear transient growth problem}\label{A}

\begin{figure}
\begin{center}
\includegraphics[trim= 4cm 16.5cm 4cm 4cm,clip=true,width=14cm, height=10cm]{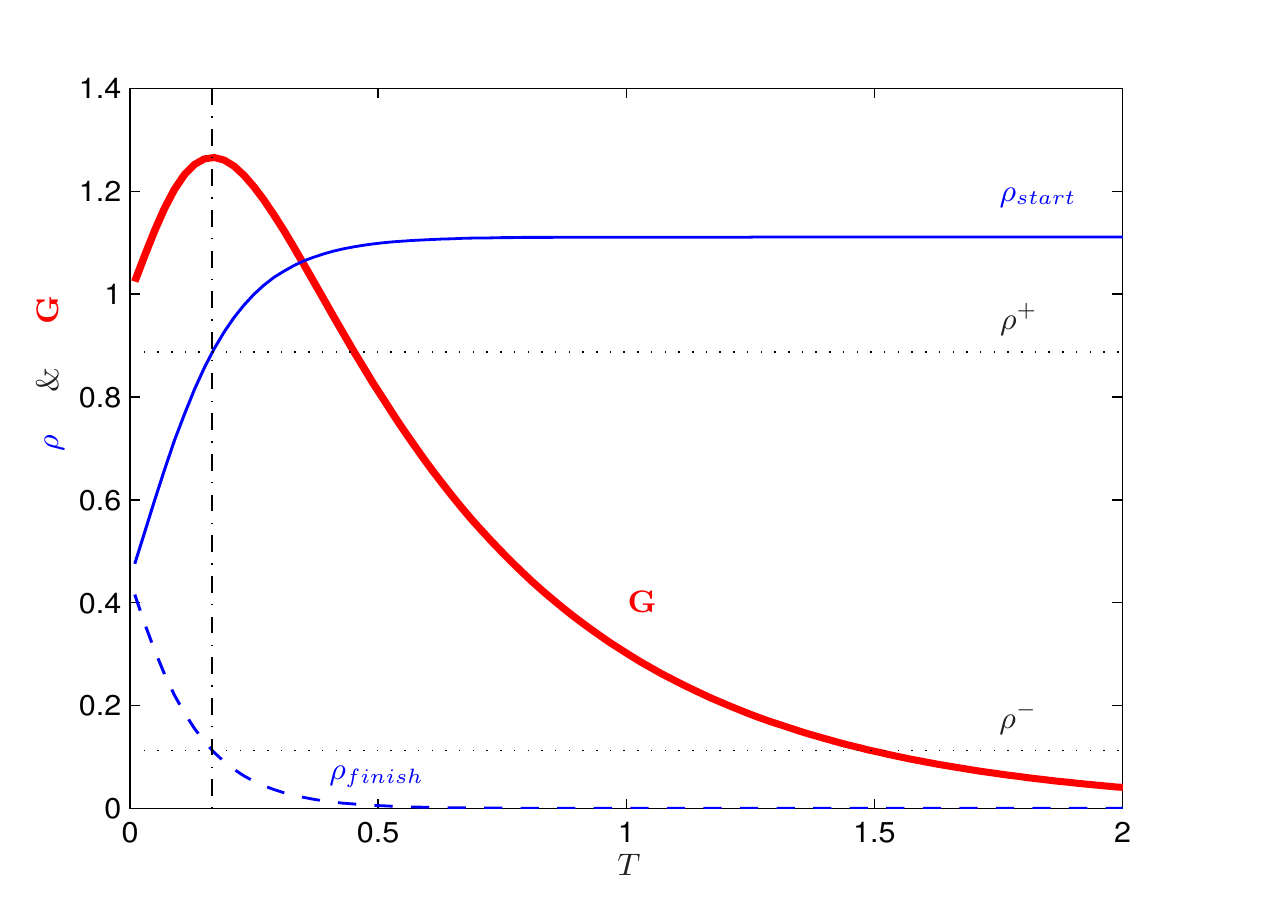}
\end{center}
\caption{The maximum gain $G$ (thick red line) against target time $T$
  for the linear system (\ref{linear_problem}). The thinner
  solid/dashed blue lines indicate the starting/finishing values of
  $\rho:=x_2/x_1$ for the corresponding optimal. Also shown as
  horizontal (black) dotted lines are the orientations where energy
  growth starts ($\rho^+$) and finishes ($\rho^-$ ) ($\rho$ decreases
  in time for $x_2(0)>0$). Notice that the maximum growth occurs
  precisely when the initial condition has $\rho(0)=\rho^+$ {\em and}
  $T$ is such that $\rho(T)=\rho^-$.}
\label{GvsT}
\end{figure}

The dynamics in the simple ODE model (\ref{ODE}) around the equilibrium $\bX_0=(0,0)$ are linearised when $d \ll 1$.
In this limit, the equation for $\bx$ is just  
\beq
\frac{d \bx}{dt} = L \bx,\quad {\rm where} \quad L:=
\left[ \begin{array}{rr} -1 & 10 \\ 0 & -10 \end{array} \right]
\label{linear_problem}
\eeq
and standard matrix manipulations can then calculate the maximal achievable distance after a time $T$. 
$L$ has eigenvalues $\lambda_1=-1$ and $\lambda_2=-10$ and corresponding
eigenvectors $\mathbf{q}_1$ and $\mathbf{q}_2$. Defining the matrix ${\bf Q}:=(\mathbf{q}_1 | \mathbf{q}_2 )$ (the 2 eigenvectors arranged in columns) and 
\beq 
e^{ \mathbf{\Lambda} T}= 
\left[ \begin{array}{cc}
e^{\lambda_1 T} &        0      \\ 
0               & e^{\lambda_2T} 
         \end{array} \right]
\eeq
then, if ${\bf a}$ is a vector specifying the initial condition
\beq
\bx(0)=\sum^{2}_{j=1} a_j \mathbf{q}_j={\bf Q} {\bf a}
 \quad 
\Rightarrow \quad \bx(t)=\sum^{2}_{j=1} a_j e^{\lambda_j T} \mathbf{q}_j
=
{\bf Q} e^{\mathbf{\Lambda} T} {\bf a}.
\eeq
Then $\max {\cal L}:=G\, d^2 $ where  the gain 
\beq
G(T)=\max_{\mathbf{a}} \frac{\|\bx(T)\|_2^2 }
                               {\|\bx(0)\|_2^2 }
=\max_{\mathbf{a}} 
\frac{{\bf a}^\dag e^{\mathbf{\Lambda} T}{\bf Q}^\dag       {\bf Q} e^{\mathbf{\Lambda} T} {\bf a} }
     {{\bf a}^\dag {\bf Q}^\dag {\bf Q} {\bf a} }=\| {\bf M} \|_2^2
\eeq
and ${\bf M}:= {\bf Q} e^{\mathbf{\Lambda} T} {\bf Q}^{-1}$ ($^\dag$
indicating transpose). $G$ is therefore the largest singular value of
${\bf M}$ or equivalently the largest (real) eigenvalue of the
symmetric matrix ${\bf M}^\dag {\bf M}$. The linear optimal
$\theta=\theta^*$ increases monotonically from 0.1413 ($25.4^o$) at
$T=0$, through 0.2322 ($41.8^o$) at T=0.16615 where $G$ peaks at
1.26590, and then onto $0.2667$ ($48.0^o$) where $G \rightarrow 0 $ as
$T \rightarrow \infty$: see Figure \ref{GvsT}.

%
%

In 2D, the situation is sufficiently simple to analyse completely for the general linear problem
\beq
\frac{d \bx}{dt} = L \bx:= \left[ \begin{array}{rr} -a & b \\ 0 & -c \end{array} \right]
\bx
\label{linear_problemII}
\eeq 
with $a,b$ and c all positive real numbers (the interesting
stable case). Defining $\rho(t)$ as the orientation $x_2(t)/x_1(t)$
and energy $E(t):=\half \bx(t)^2$, energy growth begins at the
orientation $\rho^+$ and ends at the orientation $\rho^-$ where
\beq
\rho^{\pm}:= \frac{b \pm \sqrt{b^2-4ac}}{2c}.
\eeq
The eigenvalues of $L$ are $-a$ and $-c$ and with their corresponding eigenvectors, the general solution to (\ref{linear_problemII}) can be written down as
\beq
\bx(t)= \alpha \left[ \begin{array}{c} 1       \\ 0 \end{array} \right] e^{-at}
        +\beta \left[ \begin{array}{c} b/(a-c) \\ 1 \end{array} \right] e^{-ct}. 
\eeq
Imposing the conditions that $\rho(0)=\rho^+$ and $\rho(T)=\rho^-$ requires
\beq
\frac{\beta}{\alpha}=\frac{\rho^+(c-a)}{c-a+b \rho^+}, \qquad
T^*:= \frac{1}{c-a} \log 
\biggl[
\left(\frac{\rho^+}{\rho^-} \right)
\frac{c-a+b \rho^-}
     {c-a+b \rho^+}
\biggr]
\label{T}
\eeq
so that
\beq
G_{\max}=\frac{\rho^{+2}(\rho^{-2}+1)}
              {\rho^{-2}(\rho^{+2}+1)} e^{-2cT^*}.
\label{Gmax}							
\eeq							

\section*{Appendix B: Linear verses nonlinear transient growth}\label{B}

\begin{figure}
\begin{center}
\includegraphics[trim= 3.5cm 9cm 3.5cm 9cm,clip=true,width=13.5cm, height=11cm]{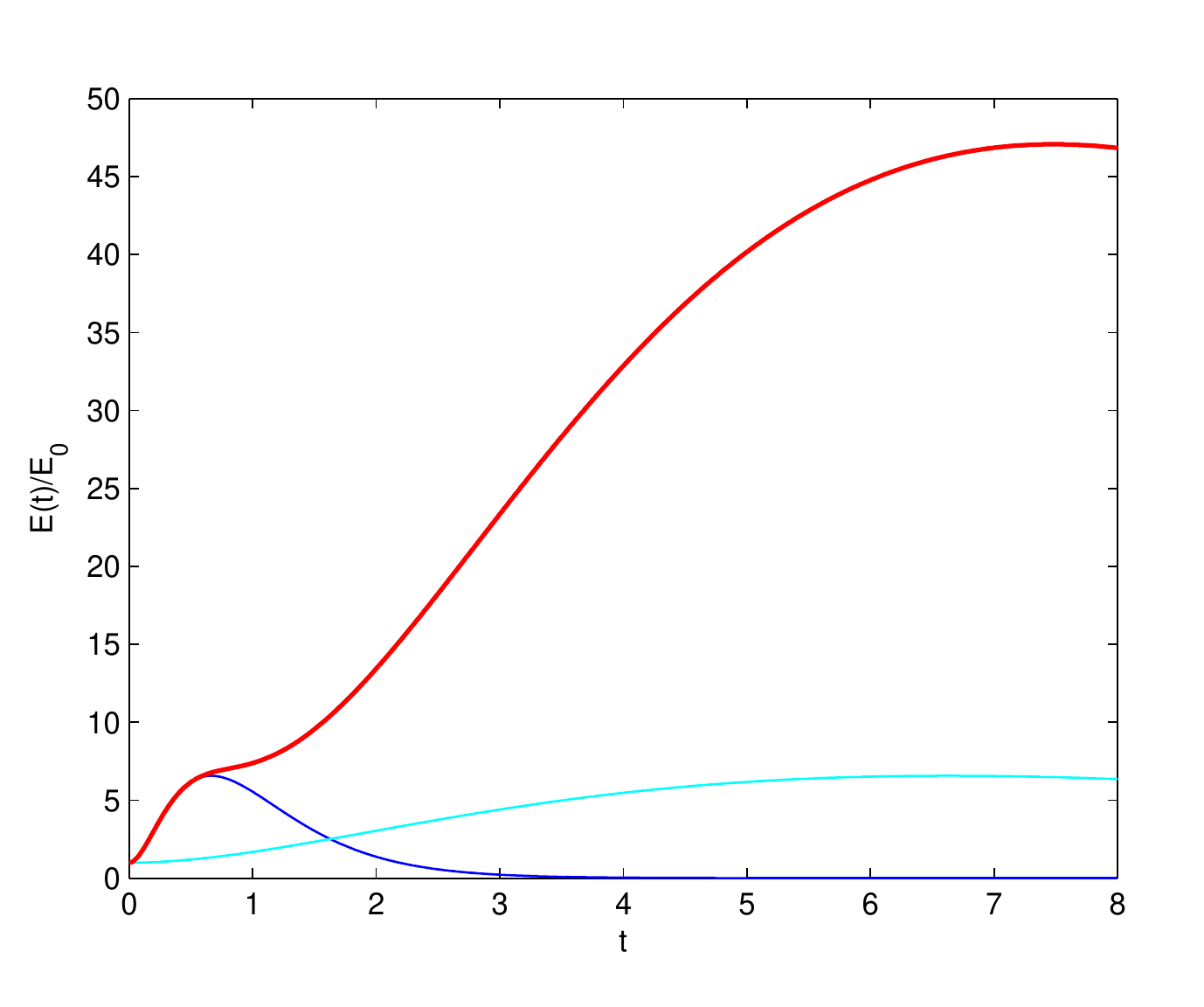}
\end{center}
\caption{Linear and nonlinear transient growth for the system
  (\ref{linear_problemIII}) with $(a,b,c,\eps)=(1,10,2,0.1)$.  The
  thin blue and cyan lines show optimal growth in $U_1$
  and $U_2$ which both have a peak gain $G$ of
  6.57 but at different times $T^*=0.66$ and $T^*/\eps$ respectively
  for infinitesimally small initial energies. The thick red line is
  the result of using the optimal initial condition (\ref{ic}) for the
  $U_1$ with $\delta=0.15$ which is sufficient to
  allow the two transient growth processes to combine to produce much
  larger overall growth. Note the similarity of this nonlinear curve
  to that in Figure \ref{Lin_Nonlin}.}
\label{Gvst}
\end{figure}

Here we discuss how nonlinear optimals are related to linear optimals
when the nonlinearity in the system preserves the functional being
optimised (a prime example being the energy-preserving nonlinearity of
the Navier-Stokes equations in fluid mechanics). Consider an ODE
system whose phase space is partitioned into subspaces each of which
is invariant under the linearised dynamics about the origin $\bx={\bf
  0}$ which is a stable fixed point. If some of the subspaces support
transient growth then it is possible for nonlinearities in the system
to couple these growth processes to give much larger overall growth
than any possible in the linearised dynamics. To illustrate this, we
take a simple 4D system which has two such subspaces, 
\beq 
\dot{x}:=
\frac{d \bx}{dt} = \left[ \begin{array}{rrrr} -a & b & 0 & 0 \\ 0 & -c
    & 0 & 0 \\ 0 & 0 & -\eps a & \eps b \\ 0 & 0 & 0 & -\eps c \\
\end{array} \right]
\bx+ 
\left[\begin{array}{c}
-x_1 x_4\\0\\0\\ x_1^2
\end{array}\right].
\label{linear_problemIII}
\eeq
For simplicity, the linear dynamics in each subspace are identical up
a change in timescale which means that the same maximum transient
growth (~$G_{\max}$ given by (\ref{Gmax})~) occurs in each but at two
different times (given by (\ref{T})\,); $T^*$ for subspace
$U_1:=\{(x_1,x_2,0,0)\,|\, x_1,x_2 \in {\mathbb R} \}$ and $T^*/\eps$
for subspace $U_2:= \{(0,0,x_3,x_4) \,|\, x_3,x_4 \in {\mathbb R}\}$:
see Figure \ref{Gvst}.  Minimal nonlinear terms are included
designed to a) conserve energy (as in the Navier-Stokes equations) and
b) to allow the faster energy growth in subspace $U_1$ to pump-prime
the slower growth in subspace $U_2$. This is clearly seen to occur for
$\delta$ large enough when taking the optimal initial condition for
$U_1$ 
\beq 
\bx(0)=(x_1,x_2,x_3,x_4)=\delta (1,\rho^+,0,0),
\label{ic}
\eeq
see Figure \ref{Gvst} for an example using $(a,b,c,\eps)=(1,10,2,0.1)$ 
(for the linear problem in $U_1$, $G_{\max}=6.57$ and $T^*=0.66$).

%
%

If the terms `LOP' and `NLOP' indicate the global linear and nonlinear
optimals over asymptotically large $T$, both clearly must have 
\beq
\dot{E}(0)=0
\eeq 
($E:=\half \bx^2$) otherwise the time origin could be adjusted to
increase the overall growth \cite{C05}. Since $\dot{E}$ is only
determined by the linear terms (the nonlinear terms can't contribute
as they are energy-preserving), then all the candidate initial
conditions satisfying the constraint $\dot{E}=0$ for the full
nonlinear optimisation problem are actually present in the linear
problem. However, the linear problem may never select the NLOP for
{\em any} $T$ because without nonlinearity its growth is overshadowed
by other candidates (i.e. the NLOP may only be a local rather than
global maximum or not even a maximum at all!). To illustrate this with
a concrete example, consider a 6D extension of the above 4D system by
adding another 2D subspace $U_3$ to $U_1 \bigoplus U_2$. If the linear
transient growth in $U_3$ always produces slightly more growth over
any choice of $T$ than $U_1$ (at least until the slower growth in
$U_2$ takes over), the linear optimisation problem will never select
any candidate initial condition which has some projection in $U_1$. In
contrast the NLOP will be the linear optimal for $U_1$
(i.e. (\ref{ic})\,) since this is the only way $\dot{E}(0)$ vanishes
non-trivially in $U_1$. This situation is borne out by the NLOP found
in (very high dimensional) pipe flow \cite{PK10,PWK12}. In just 2D,
however, there are only two candidates in the linear problem which are
joint global optimisers. Hence in this case the NLOP has to be
contained in the linear problem trivially.


\end{document}